\documentclass[manuscript]{aastex}
\usepackage{graphicx}
\usepackage{bm}
\usepackage{subfigure}

\let\la=\langle
\let\ra=\rangle
\shorttitle{Spherical transfer}
\shortauthors{Anusha, Nagendra, Paletou, L\'eger}
\begin{document}
\title{Preconditioned Bi-Conjugate Gradient Method for Radiative Transfer in Spherical Media}
\author{L.~S. Anusha and K.~N.~Nagendra}
\affil{Indian Institute of Astrophysics, Koramangala,
Bangalore 560 034, India}
%\email{anusha@iiap.res.in}
%\email{knn@iiap.res.in}
\and
\author{F.~Paletou and L.~L\'eger}
\affil{Laboratoire d'Astrophysique de Toulouse-Tarbes, Universit\'e de Toulouse, CNRS,
14 av. E. Belin, 31400 Toulouse, France.}
\begin{abstract}
A robust numerical method called the Preconditioned Bi-Conjugate Gradient (Pre-BiCG)
method is proposed for the solution of radiative transfer equation in spherical geometry. 
A variant of this method called Stabilized Preconditioned Bi-Conjugate Gradient
(Pre-BiCG-STAB) is also presented. These are iterative methods based on the construction 
of a set of bi-orthogonal vectors. The application of Pre-BiCG method in some 
benchmark tests show that the method is quite versatile, and can handle hard problems that
may arise in astrophysical radiative transfer theory.
\end{abstract}
\section{Introduction}
\label{intro}
The solution of transfer equation in spherical geometry remains a classic problem
even 75 years after the first attempts by \citet{cha34, kos34}, 
who used Eddington approximation. In later decades more accurate methods 
were given \citep[see][for historical reviews]{mih78, per02}. \citet{hum71}, 
\citet{kun73, kun74} developed a variable Eddington factor method, and computed 
the solution on rays of constant impact parameter  
(tangents to the discrete shells and parallel to line of sight) in 1D spherical geometry. 
This is a very efficient differential equation based technique which uses Feautrier solution 
along rays of constant impact parameter.
An integral equation method was developed by 
\citet{sch74} to solve the problem, again based on tangent rays approach. 
\citet{per73} presented a highly accurate 
finite difference method based on the first order form of the transfer equation.
All these methods were later extended to expanding, and 
highly extended atmospheres. However, in this paper we confine our attention to static, 
1D spherical atmospheres.

In a next epoch in the development of spherical radiative transfer, the integral operator
techniques were proposed. The idea of operator splitting and the use of approximate operators
in iterative methods was brought to the astrophysical radiative transfer 
in planar media by \citet{can73}. \citet{sch81} extended his work with a new
definition of the approximate operator. The application of integral operator 
technique to the spherical transfer started with the work of 
\citet{hum85} and \citet{wer85}. They used approximate operators
that are diagonal, constructed from core saturation approach.
The $\hat \Lambda$ operator contains the non-local
coupling between all the spatial points. \citet{ols86} 
showed that the diagonal part (local coupling) of the actual 
$\hat \Lambda$ operator itself is an optimum choice for the 
`approximate operator'. These methods are known as approximate
Lambda Iteration (ALI) methods. The ALI methods which are based on the 
concept of operator splitting and the use of Jacobi iterative technique, were 
widely used in the later decades in radiative transfer theory 
\citep[see][for historical reviews]{hub03, hum03}.

\citet{gro97} used an implicit integral 
method to solve static spherical line transfer
problems. The most recent and interesting work on spherical radiative transfer are the 
papers by \citet{aar06} and \citet{dan08} both of which are based on 
Gauss-Seidel (GS) and Successive Over Relaxation (SOR) iterative techniques.

\citet{kle89} were the first to use BiCG technique in astrophysics.
They use BiCG with incomplete LU decomposition technique in their double
splitting iterative scheme along with Orthomin acceleration. They applied it
to multi-dimensional line transfer problem. \citet{aue91} describes a variant
of Orthomin acceleration which uses `Minimization with respect to a set of
Conjugate vectors'. He uses a set of n (usually n=2 or n=4) conjugate
direction vectors which are orthogonal to each other, constructed using
the residual vectors with a purpose to accelerate the convergence sequence.

\citet{hub07} developed GMRES (actually its variant called Generalized
Conjugate Residuals GCR) method to solve the spherical transfer problem. It
is based on an application of the idea of Krylov subspace techniques. They applied
it to a more general time-dependent transport with velocity fields in a medium
which scatters anisotropically. They apply GMRES method to the neurino transfer.
It can also be used for radiation transfer problem, including the simple
problem of 2-level atom line transfer discussed in this paper.

The Preconditioned Bi-Conjugate Gradient method \citep[hereafter 
Pre-BiCG, see eg.,][]{saad00} was  first introduced to the line transfer
in planar media, by \citet[][]{pal09} who describe the method and compare
it with other prevalent iterative methods, namely GS/SOR.
In this paper we adopt the Pre-BiCG method 
to the case of spherical media. We also show that the 
`Stabilized Preconditioned Bi-Conjugate Gradient (Pre-BiCG-STAB)' is
even more advantageous in terms of memory requirements but with similar
convergence rate as Pre-BiCG method.

It is well known that the spherical radiative
transfer in highly extended systems, despite being a straight forward
problem, has two inherent numerical difficulties namely (i) peaking of the radiation
field towards the radial direction, and (ii) the $(1/r^2)$ dilution 
of radiation in spherical geometry.
To handle these, it becomes essential to take a very large number of angle ($\mu$) points
and spatial ($\tau$) points respectively. The existing ALI methods clearly slow down
when extreme angular and spatial resolutions are demanded (for example see table~\ref{table_1}).
Therefore there is a need to look for a method that is as efficient as ALI methods, but
faster, and is relatively less sensitive to the grid resolution.
The Pre-BiCG method and a variant of it provide such an alternative as 
we show in this paper.

Governing equations are presented in Sect.~\ref{transfer}. 
In Sect.~\ref{geo}, we define the geometry of the problem and the specific details of
griding. In Sect.~\ref{model}, the benchmark models are defined. 
We briefly recall the Jacobi, and GS/SOR methods
in Sect.~\ref{ALI}. In Sect.~\ref{bicg_sphere} 
we describe the Pre-BiCG method. The computing algorithm is
presented in Sect.~\ref{prebicg_alg}. In Sect.~\ref{prebicg-stab}
we describe the Pre-BiCG-STAB method briefly, and 
we give the computing algorithm in Sect.~\ref{stab_alg}.
In Sect.~\ref{bicg_ali} we compare the performance of Pre-BiCG with the Jacobi, and 
GS/SOR methods. In Sect.~\ref{results} we validate this new method, 
by comparing with the existing well known benchmark solutions in spherical 
line radiative transfer theory. Conclusions 
are presented in Sect.~\ref{conclusions}.

\section{Radiative transfer in a spherical medium} 
\label{method}

\subsection{The transfer equation}
\label{transfer}

In this paper we restrict ourselves to the case of a 2-level atom model.
Further, we assume complete frequency redistribution (CRD). The transfer equation 
in divergence form is written as 
\begin{eqnarray}
& &\mu \frac{\partial \bm{I}(r, \mu, x)}{\partial r}+\frac{1-\mu^2}{r}
\frac{\partial  \bm{I}(r, \mu, x)}{\partial \mu} \nonumber \\
& & = [\chi_{\rm L}(r)\phi(x)+\chi_{\rm C}(r)]\nonumber \\
& & \times [\bm{S}(x, r)-\bm{I}(r, \mu, x)].
\label{transfer_div}
\end{eqnarray}
Here, $\bm{I}$ is the
specific intensity of radiation, $\bm{S}$ - the source function, $r$ - the
radial distance, $\mu$ - the direction cosine, $x$ -
the frequency measured in Doppler width units from line center, $\phi(x)$
- the line profile function, and $\chi_{\rm L}(r)$, $\chi_{\rm C}(r)$
- line center and continuum opacities respectively. The differential optical depth 
element is given by
\begin{equation}
{\rm d}\tau(r)=-\chi_{\rm L}(r) {\rm d}r.
\label{dtau}
\end {equation}
There are several methods which use the above form of the transfer equation
\citep[see][]{per02}. In our paper we solve the transfer equation on a set of 
rays tangent to the spherical shells.
It is written as
\begin{eqnarray}
& &\pm \frac{\partial \bm{I}^{\pm}(z, p, x)}{\partial z}
=[\chi_{\rm L}(r)\phi(x)+\chi_{\rm C}(r)]  \nonumber \\
& & \times [\bm{S}(x, r)-\bm{I}^{\pm}(z, p, x)], 
\label{transfer_red}
\end{eqnarray}
for the outgoing (+) and incoming (-) rays respectively. 
Here $z$ is the distance along the tangent rays and 
$p$ is the distance from the center to the points on the vertical axis
(the mid-line), where the 
tangent rays intersect it (see Fig.~\ref{fig_geometry}). 
The direction cosines $\mu\,\,(0 \le \mu \le 1)$ 
are related to $p$ by $\mu=\sqrt {1-(p^2/r^2)}$ for a shell of radius $r$. 
The optical depth scale along the tangent rays are now computed using 
${\rm d}\tau(z)={\rm d}\tau(r)/\mu$. 
In practical work, due to the symmetry of the problem, 
it is sufficient to perform the computations on a quadrant only.
The source function is defined as 
\begin {equation}
\bm{S}(x, r)=\frac{\chi_{\rm L}(r)\phi(x)\bm{S}_{\rm L}(r)+\chi_{\rm C}(r)\bm{S}_{\rm c}(r)}
{\chi_{\rm L}(r)\phi(x)+\chi_{\rm C}(r)}.
\label{source_total}
\end{equation}
$\bm{S}_{\rm c}(r)$ is the continuum source function taken as the Planck function
$\bm{B}_{\nu}(r)$ throughout this paper.
The monochromatic optical depth scale $\Delta \tau_{x}=
\Delta \tau_{\rm z} [\phi(x)+\beta_{\rm c}]$, with 
$\beta_{\rm c}=\chi_{\rm C}(r)/\chi_{\rm L}(r)$ along
the tangent rays. For simplicity, hereafter we omit the subscript 
$z$ from $\tau_{\rm z}$ and write $\tau$ to denote $\tau_{\rm z}$. 
The line source function is given by
\begin{eqnarray}
& &\bm{S}_{\rm L}(r)=(1-\epsilon) 
\int_{-1}^{1}\frac{{\rm d}\mu'}{2} \,
\int_{-\infty}^{\infty}{\rm d}x' \, \nonumber \\
& &\phi(x') \bm{I}(\tau, \mu', x')+\epsilon \bm{B}_{\rm \nu}(r),
\label{source_line}
\end{eqnarray}
with the thermalization parameter defined in the conventional manner as
$\epsilon=C_{\rm ul}/(A_{\rm ul}+C_{\rm ul})$, where $C_{\rm ul}$ and $A_{\rm ul}$
are collisional and radiative de-excitation rates. 
The intensity along the rays is computed using the formal solution integral 
\begin{eqnarray}
& &\bm{I}^{+}(\tau, p, x)=\bm{I}^{+}_{0}(\tau, p, x) \exp[-\Delta \tau_{x}]+\nonumber \\
& & \int_{\tau}^{T} 
\exp[-\Delta \tau'_{x}] \bm{S}(\tau')[\phi(x)+\beta_{\rm c}]{\rm d}\tau' .
\label{int_plus}
\end {eqnarray}
The corresponding integral for the incoming rays is
\begin{eqnarray}
& & \bm{I}^{-}(\tau, p, x)=\bm{I}^{-}_{0}(\tau, p, x)\exp[-\Delta \tau_{x}]+\nonumber \\
& & \int_{0}^{\tau}
\exp[-\Delta \tau'_{x}] \bm{S}(\tau')[\phi(x)+\beta_{\rm c}]{\rm d}\tau'.
\label{int_minus}
\end{eqnarray}
Here, $\bm{I}^{+}_{0}(\tau, p, x)$ represents the inner boundary condition imposed at the
core and along the mid-vertical line (see Fig.~\ref{fig_geometry}). 
$\bm{I}^{-}_{0}(\tau, p, x)$ is the outer 
boundary condition specified at the surface of the spherical atmosphere.
When the above formal integral is applied to a stencil of short characteristic (MOP)
along a tangent ray, it takes a simple algebraic form
\begin {eqnarray}
& &\bm{I}^{\pm}(\tau, p, x)= \bm{I}^{\pm}_{\rm O}(\tau, p, x) \exp[-\Delta \tau_{\rm M}]+ \nonumber \\
& & \bm{\Psi}^{\pm}_{\rm M}(\tau, p, x)\bm{S}_{\rm M}+\bm{\Psi}^{\pm}_{\rm O}(\tau, p, x)\bm{S}_{\rm O}+ \nonumber \\
& & \bm{\Psi}^{\pm}_{\rm P}(\tau, p, x)\bm{S}_{\rm P} ,
\label{int_sc}
\end{eqnarray}
where $\bm{S}_{\rm M, O, P}$ are the source function values at M, O and P
points on a short characteristic. The coefficients $\bm{\Psi}$ are calculated following the
method described in \citet{kun88}.
\subsection{The constant impact parameter approach}
\label{geo}

In Fig.~\ref{fig_geometry}, we show the geometry used for computing the 
specific intensity $\bm{I}(\tau, p, x)$
along rays of constant impact parameter. 

In a spherically symmetric medium, we first discretise the radial co-ordinate $r$
($R_{\rm core} \le r \le R$), where $R_{\rm core}$ is the core radius, and $R$ is the outer radius of 
the atmosphere. The radial grid is given by $r_{\rm k}, k=1, 2, \ldots, N_{\rm d}$, where $r_{\rm 1}$ 
is the radius of the outer most shell, and $r_{\rm N_d}$ is that of the inner most shell. 
$d\Omega/{4\pi}$ is the 
probability that the direction of propagation of an emitted photon lies within an element of 
solid angle $d\Omega$. In the azimuthally symmetric case, it is $d\mu/2$. To calculate 
the mean intensity $\bm{\bar{J}}$ in plane parallel geometry, we integrate 
the intensity over the angular variable $\mu $ itself. In spherical medium, 
we have one to one correspondence between the ($\mu$,\,$r$) and the ($p$,\,$r$) system. 
In  ($p$,\,$r$) system, the probability that a photon is 
emitted with its impact parameter between $p$ and $p$+$dp$, propagating in either positive 
$\mu$ or negative $\mu$ direction is $pdp/{2r\sqrt{r^2-p^2}}$. The direction cosines 
made by the rays in the ($\mu$,\,$r$) space, with a tangent ray of constant $p$ value, 
are given by $\mu_{\rm i}=\sqrt {1-(p^2/r_{\rm i}^2)}$ at different radii $r_{\rm i}$. 
Therefore the angular integration factor $d\mu/2$ can be changed to 
$pdp/{2r\sqrt{r^2-p^2}}$ \cite[see][]{kun73}. 

{\it The $p$ - grid construction}:
If $N_{\rm c}$ is the number of core rays, then the $p$-grid for 
the core rays is computed using:
\begin{eqnarray}
& & \textrm{do}\,\,i=1,\,N_{\rm c}\nonumber \\
& & p(i)=R_{\rm core}*\left(\sqrt{1-(i/N_{\rm c})^2}\right) \nonumber \\
& & 0\  <\ p\ <\  R_{\rm core}       \nonumber \\
& & \textrm{end do}\nonumber
\label{p_grid}
\end{eqnarray}
The number of lobe rays equals the number of radial points.
For lobe rays, the $p$-grid is same as radial $r$-grid. It is constructed using:
\begin{eqnarray}
& & \textrm {do}\,\,i=1,\,N_{\rm d} \nonumber \\
& & p(N_{\rm c}+i)=r(i) \nonumber \\
& &\textrm {end do} \nonumber
\label{pc_grid}
\end {eqnarray}
\\where 
$N_{\rm d}$=the number of radial points.
Thus, the total number of impact parameters is
$N_{\rm p}=N_{\rm c}+N_{\rm d}$. We have followed \citet[][]{aue84}
in defining the $p$-grid in this manner.

\subsection{Benchmark models}
\label{model}
Geometrical distances along the rays of constant impact parameter are constructed as follows:
\begin{equation}
z(p, r)=\sqrt{r^2 - p^2} .
\label{dist_geo}
\end{equation}
For spherical shells we perform several tests using power-law type variation of
density. For such atmospheres, the line and continuum opacities also vary as 
a power law given by
\begin{equation}
\chi_{\rm L, C}(r) \propto r^{-\tilde n}.
\label{chi_lc}
\end{equation}
Let $C$ and $\bar C$ denote the proportionality constants for $\chi_{\rm L}(r)$
and $\chi_{\rm C}(r)$ respectively. The constant $C$ can be determined 
using the optical depth at line center $T$. For a power law with index $\tilde n$,
\begin{equation}
C=\frac{T(1-\tilde n)}{{R^{(1-\tilde n)}}-{R_{\rm core}^{(1-\tilde n)}}}.
\label{chil_const}
\end{equation}
Using the given input value of $\beta_{\rm c}={\bar C}/C$ we can compute the 
constant ${\bar C}$. 

We use Voigt profile with damping parameter $a$ or the Doppler profile for the
results presented in this paper.
The spherical shell atmosphere is characterized by the following parameters:
($R$, $\tilde n$, $T$, $a$, $\epsilon$, $\beta_{\rm c}$, $B_{\rm \nu}$).
We recall that $R$ is the outer radius of the spherical atmosphere surrounding a
hollow central cavity of radius $R_{\rm core}$. When $R=R_{\rm core}$ we recover the 
plane parallel limit. For the spherical shell atmospheres, we take $R_{\rm core}=1$
as the unit of length to express the radial co-ordinate.
The boundary conditions are specified at the outer boundary 
($ \bm{I}^{-}(\tau=0, p, x)=0$) and
the inner boundary. There are two types of inner boundary conditions:\\

\noindent
(a) {\it Emitting Core}:\\
\noindent
Core rays: 
\begin{equation}
\bm{I}^{+}(\tau=T, p \le R_{\rm core}, x)=B_{\rm \nu}.
\end{equation}

\noindent
Lobe rays: 
\begin{equation}
\bm{I}^{+}(\tau=T, p=r_{\rm i}, x)=\bm{I}^{-}(\tau=T, p=r_{\rm i}, x),
\end{equation}
$i=1, 2, \ldots, N_{\rm d}$ along the mid-vertical.\\

\noindent
(b) {\it Hollow Core: }\\
For both the core and the lobe rays:
\begin{equation}
\bm{I}^{+}(\tau=T, p, x)=\bm{I}^{-}(\tau=T, p, x).
\end{equation}
The hollow core boundary condition is also called `planetary nebula boundary
condition' \citep[see][]{mih78}. It is clear that a spherical shell with a
hollow core is equivalent to a plane parallel slab of optical 
thickness $2T$ with symmetry about the mid-plane at $\tau=T$. 
We use spherical shell atmospheres for most of our studies.

\subsection{Iterative methods of ALI type for a spherical medium}
\label{ALI}
The ALI methods have been successfully used for the solution of transfer equation in
spherical shell atmospheres \citep[see eg.,][and references therein]{hum03}. 
These authors use the Jacobi iterative methods (first introduced
by \citet{ols86}) for computing the source function 
corrections. Recently the GS method has been proposed 
to solve the same problem \citep[see eg.,][]{aar06, dan08}. 
\citet{hub07} proposed the GMRES method for
solving spherical radiative transfer problem. GMRES and
Pre-BiCG both belong to Krylov subspace technique.
In this paper we compute the spherical
transfer solutions by Jacobi and GS/SOR methods, and compare with the solutions computed using
the Pre-BiCG method. For the sake of clarity, 
we recall briefly the steps of Jacobi and GS/SOR methods.\\

\noindent
{\it Jacobi Iteration Cycle}: The source function corrections are given by
\begin {equation}
\delta \bm{S}_{\rm k}^{n}=\bm{S}_{\rm k}^{n+1}-\bm{S}_{\rm k}^{n}=
\frac{(1-\epsilon)\bm{\bar{J}}_{\rm k}^{n}
+\epsilon \bm{B}_{\nu}-\bm{S}_{\rm k}^{n}}
{[1-(1-\epsilon) \hat \Lambda^{*}_{\rm k,k}]},
\label{deltas_jacobi}
\end {equation}
for the $n$th iterate. Here ${\rm k}$ is the depth index. The $\hat \Lambda^{*}$ is the approximate
operator which is simply taken as the diagonal of the actual  $\hat \Lambda$ operator defined
through 
\begin{equation}
{\hat \Lambda}[\bm{S}]=\bm{\bar{J}};
\label{j_bar}
\end{equation}
\begin{equation}
\bm{\bar{J}}(\tau)=\int_{-1}^{+1}\frac{{\rm d}\mu'}{2}\,\int_{-\infty}^{\infty}{\rm d}x'\, 
\phi(x')\bm{I}(\tau, \mu', x').
\label{jbar}
\end{equation}

\noindent
{\it GS/SOR Iteration Cycle}: The essential difference between the Jacobi and GS/SOR 
methods is the following: 
\begin {equation}
\bm{S}_{\rm k}^{n+1}=\bm{S}_{\rm k}^{n}+\omega\,\delta \bm{S}_{\rm k}^{n}. \nonumber \\
\end{equation}
Here the parameter $\omega$ is called the relaxation parameter which
is unity for the GS technique.

The SOR method is derived from the GS method by simply taking
$1 < \omega < 2$ \citep[see][for details]{tru95}. 
The source function correction for the GS method is given by
\begin{equation}
\delta \bm{S}_{\rm k}^{n} =\frac{(1-\epsilon)\bm{\bar{J
}}_{\rm k}^{n(old+new)}+\epsilon \bm{B}_{\nu}-\bm{S}_{\rm k}^{n}}
{[1-(1-\epsilon) \hat \Lambda^{*}_{\rm k, k}]},
\label{deltas_gs}
\end {equation}
where the quantity $\bm{\bar{J}}_{\rm k}^{n(old+new)}$ denotes the mean intensity computed using new
values of the source function as soon as they become available. For those depth points for which 
the source function correction is not yet complete, GS method uses the values of the source function
corresponding to the previous iteration \citep[see][]{tru95}. For clarity we explain how the GS
algorithm works in spherical geometry, on rays of constant impact parameter.\\

\noindent
Begin loop over iterations\\
Begin loop over radial shells with index $k$  \\
Begin loop over impact parameters (or directions) with increasing $p$ \\

\noindent
For the $n$th iteration:

\noindent
\begin{center}{\bf For the incoming rays ($\mu < 0$):} \\
({\bf Reverse sweep along radial shells})
\end{center}

\noindent
(a) This part of the calculations start at the outer 
boundary for all impact parameter rays. \\

\noindent
(b) $\bm{I}_{\rm k}$ are first calculated for a given radial shell $k$
using $\bm{S}^{\rm n}_{\rm k}$, $\bm{S}^{\rm n}_{\rm k-1}$ and 
$\bm{S}^{\rm n}_{\rm k+1}$. \\

\noindent
(c) The partial integral $\bar{\bm{J}}_{\rm k}(\mu<0)$ are calculated
 before proceeding to the
next shell. This part of the calculations is stopped
when the core (for the core rays) 
and the mid-vertical line (for the lobe rays) are reached. \\

\noindent
\begin{center}{\bf For outgoing rays ($\mu > 0$):} \\
({\bf Forward sweep along radial shells})
\end{center}

\noindent
(d) This part of the calculations start at the inner boundary.
First, for the radial shell with $k=N_{\rm d}$ \\
$\bar{\bm{J}}_{\rm N_{\rm d}}$ is calculated, using boundary conditions
$\bm{I}_{\rm N_{\rm d}}$.\\

\noindent
(e) $\delta \bm{S}^{\rm n}_{\rm N_{\rm d}}$ is computed and
the source function is updated using 
$\bm{S}^{\rm n+1}_{\rm N_{\rm d}}=\bm{S}^{\rm n}_{\rm N_{\rm d}}
+\delta \bm{S}^{\rm n}_{\rm N_{\rm d}}$.\\

\noindent
(f) For the next radial shell $k=N_{\rm d}-1$: \\
to calculate $\bm{I}_{\rm N_{\rm d}-1}$
by applying the short characteristic formula,
$\bm{S}_{\rm N_{\rm d}}$, $\bm{S}_{\rm N_{\rm d}-1}$ and $\bm{S}_{\rm N_{\rm d}-2}$
are needed. Already $\bm{S}^{\rm n+1}_{\rm N_{\rm d}}$, 
$\bm{S}^{\rm n}_{\rm N_{\rm d}-1}$
and $\bm{S}^{\rm n}_{\rm N_{\rm d}-2}$ are available.
GS takes advantage of the available new source function at $k=N_{\rm d}$.
$\bm{I}_{\rm N_{\rm d}-1}$ is calculated with this set of source functions.\\

\noindent
(g) Then $\bar{\bm{J}}_{\rm N_{\rm d}-1}(\mu > 0)$ are calculated using 
$\bm{I}_{\rm N_{\rm d}-1}$. \\

\noindent
(h) Note that, $\bar{\bm{J}}_{\rm N_{\rm d}-1}(\mu < 0)$ was calculated using 
$\bm{S}^{\rm n}_{\rm N_{\rm d}}$, $\bm{S}^{\rm n}_{\rm N_{\rm d}-1}$, and 
$\bm{S}^{\rm n}_{\rm N_{\rm d}-2}$ whereas $\bar{\bm{J}}_{\rm N_{\rm d}-1}(\mu > 0)$ 
used the ``updated'' source function $\bm{S}^{\rm n+1}_{\rm N_{\rm d}}$. 
Therefore $\bar{\bm{J}}_{\rm N_{\rm d}-1}$ is corrected
by adding the following correction:
\begin{eqnarray}
\Delta \bar{\bm{J}}_{\rm N_{\rm d}-1}=\delta \bm{S}^{\rm n}_{\rm N_{\rm d}} \int_{-1}^{0} 
\bm{\Psi}_{N_{\rm d}}(\mu < 0)\,\rm{d}\mu. \nonumber
\end{eqnarray}

\noindent
(i) $\delta \bm{S}^{\rm n}_{\rm N_{\rm d}-1}$ and $\bm{S}^{\rm n+1}_{\rm N_{\rm d}-1}
=\bm{S}^{\rm n}_{\rm N_{\rm d}-1}+\delta \bm{S}^{\rm n}_{\rm N_{\rm d}-1}$ are
now calculated.\\

\noindent
(j) Since ``updated'' $\bm{S}^{\rm n+1}_{\rm N_{\rm d}-1}$ at $k=N_{\rm d}-1$ is also available now,
before going to the next radial shell it is appropriate to correct the 
intensity at the present radial shell by adding to it, the following correction term
\begin{eqnarray}
\Delta \bm{I}_{\rm N_{\rm d}-1}(\mu)= \delta \bm{S}^{\rm n}_{\rm N_{\rm d}-1} 
\bm{\Psi}_{\rm N_{\rm d}-1}. \nonumber
\end{eqnarray}

\noindent
End loop over impact parameters (or directions)\\
End loop over radial shells\\
End loop over iterations.\\

\section{Preconditioned BiCG method for a spherical medium}
\label{bicg_sphere}
In this section we first describe the essential ideas of the Pre-BiCG method. 
The complete theory of the method is described in \citet{saad00}. 
We recall that the 2-level atom source function with a
background continuum is given by
\begin{equation}
\bm{S}(x, r)=\frac{\chi_{\rm L}(r)\phi(x)\bm{S}_{\rm L}(r)+\chi_{\rm C}(r)\bm{S}_{\rm c}(r)}
{\chi_{\rm L}(r)\phi(x)+\chi_{\rm C}(r)}.
\label{souce1_total}
\end{equation}
It can be re-written as
\begin{equation}
\bm{S}(x, r)=\tilde p(x, r)\bm{S}_{\rm L}(r)+(1-\tilde p(x, r))\bm{S}_{\rm c}(r),
\label{source_px}
\end{equation}
where 
\begin{equation}
\tilde p(x, r)=\frac{\chi_{\rm L}(r)\phi(x)}{\chi_{\rm L}(r)\phi(x)+\chi_{\rm C}(r)}.
\label{pxr}
\end{equation}
From equations~(\ref{source_line}), (\ref{j_bar}) and (\ref{jbar}), we get
\begin{eqnarray}
& & \bm{S}(x, r)=\tilde p(x, r)\{(1-\epsilon) \hat \Lambda[\bm{S}(x, r)]
+\epsilon \bm{B}_{\nu}(r)\}+ \nonumber \\
& & (1-\tilde p(x, r))\bm{S}_{\rm c}(r).
\label{source_total1}
\end{eqnarray}
Therefore the system of equations to be solved becomes
\begin {eqnarray}
& & [\hat{I}-(1-\epsilon) \tilde p(x, r) {\hat \Lambda}]\bm{S}(x, r)=
\tilde p(x, r) \epsilon \bm{B}_{\nu}(r)+ \nonumber \\
& & (1-\tilde p(x, r))\bm{S}_{\rm c}(r),
\label{system}
\end {eqnarray}
which can be expressed in a symbolic form as
\begin {eqnarray}
&\hat A \bm{y} = \bm{b}; \quad \textrm {with} \nonumber \\
&\hat A =[\hat{I}-(1-\epsilon) \tilde p(x, r) {\hat \Lambda}];\,\bm{y}=\bm{S}(x, r).
\label{system_gen}
\end {eqnarray}
The vector $\bm{b}$ represents quantities on the RHS of equation~(\ref{system}).
Now we describe briefly, how the Pre-BiCG method differs from ALI based methods.

Let $\mathbb{R}^n$ denote the $n$-dimensional Euclidean space of real numbers.

{\bf Definition}:
The Pre-BiCG algorithm is a process involving projections onto
the $m$-dimensional subspace ($m \le n$) of $\mathbb{R}^n$ 
\begin{equation}
\bm{\mathcal{K}}_{\rm m}=\textrm{span}\{\bm{v}_1, \hat A \bm{v}_1, 
\ldots, \hat A^{m-1}\bm{v}_1\},
\label{K_m}
\end{equation}
and also being orthogonal to another $m$-dimensional subspace of $\mathbb{R}^n$
\begin{equation}
\bm{\mathcal{L}}_{\rm m}=\textrm{span}\{\bm{w}_1, \hat A^{T} \bm{w}_1, \ldots, \hat A^{T(m-1)}\bm{w}_1\}.
\label{L_m}
\end{equation}
Here $\bm{v}_1$ is taken as the initial residual vector $\bm{r}_0=\bm{b}-\hat A \bm{y}_0$
with $\bm{y}_0$ the initial guess for the solution of equation ~(\ref{system_gen}).
The vector $\bm{w}_1$ is taken as arbitrary such that the 
inner product $\la \bm{v}_1, \bm{w}_1 \ra \neq 0$.
The method recursively constructs a pair of bi-orthogonal bases $\{ \bm{v}_{\rm i}; i=1, 2 \ldots, m\}$
and $\{\bm{w}_{\rm i}; i=1, 2 \ldots, m\}$ for $\bm{\mathcal{K}}_{\rm m}$
and $\bm{\mathcal{L}}_{\rm m}$ respectively, such that they satisfy the bi-orthogonality 
condition $\la \bm{v}_{\rm i}, \bm{w}_{\rm j} \ra=\delta_{\rm ij}$. 
For the purpose of application to the radiative transfer theory it is
convenient to write the Pre-BiCG steps in the form of an algorithm For 
simplicity we drop the explicit dependence on variables.

\subsection{The Preconditioned BiCG Algorithm}
\label{prebicg_alg}
Our goal is to solve equation ~(\ref{system_gen}).
In this section the symbols  $\bm{r}_{\rm i}$ and $\bm{p}_{\rm i}$
are used to be in conformity with the standard notation of residual and
conjugate direction vectors. They should not be confused with the radius
vector $r_{\rm i}$ and impact parameter $p_{\rm i}$ which appear
in spherical radiative transfer theory.

\noindent
(a) The very first step is to construct and store the matrix $\hat{A}^T$
(which does not change with iterations, for the cases considered here, namely
2-level atom model). Details of computing $\hat{A}^T$ efficiently is described
in appendix \ref{appendix_lambda}.

We follow the preconditioned version of the BiCG method.
Preconditioning is a process in which the original 
system of equations is transformed into a new system, which has faster rate of convergence.
For example, this can be done by solving the new system 
$\hat{M}^{-1} \hat{A} \bm{y}=\hat{M}^{-1} \bm{b}$ where $\hat M$ is an appropriately chosen
matrix, called the ``preconditioner'' \citep[See also equation 2 of][]{aue91}. 
This preconditioner is chosen
in such a way that,\\
(i) the new system should be easier to solve, \\
(ii) $\hat{M}^{-1}$ itself should be inexpensive to operate on an arbitrary vector,\\
(iii) the preconditioning is expected to increase the convergence rate.\\
The choice of the preconditioner depends on the problem at hand. 
When an appropriate $\hat \Lambda^*$ is chosen such that the amplification matrix
$[\hat{I}-(1-\epsilon)\hat \Lambda^*]$ has as small a maximum eigen value 
as possible \citep[see][]{ols86}, the convergence rate is enhanced. What 
enables the convergence of ALI, that satisfies the above property, and 
simplest to manipulate, is the diagonal of the $\hat \Lambda$ itself. 
Therefore the amplification matrix $[\hat{I}-(1-\epsilon) \hat \Lambda^*]$ 
with a diagonal form for $\hat \Lambda^*$ is a simple and natural choice 
as a `preconditioner'. We construct the preconditioner matrix $\hat M$ 
by taking it as the diagonal of $\hat A$.\\

\noindent
(b) An initial guess for the source function is
\begin{equation}
\bm{y}_0=\tilde{p}\, \epsilon \bm{B}+(1-\tilde p)\bm{S}_{\rm c},
\end{equation}
where the thermal part $\epsilon \bm B$ is taken as an initial guess for $\bm{S}_{\rm L}$.\\

\noindent
(c) The formal solver is used with $\bm{y}_0$ as input to calculate
$\bm{\bar J}(\bm{y}_0)$.\\

\noindent
(d) The initial residual vector is computed using
\begin {equation}
\bm{r}_0=\bm {b}-\hat{A} \bm {y}_0.\\
\label{res_init}\nonumber 
\end{equation}

\noindent
(e) The initial bi-orthogonal counterpart $\bm{r}_0^*$ for $\bm{r}_0$
is chosen such that we have $\la \bm{r}_0, \bm{r}_0^*\ra \neq 0$. One can choose 
$\bm{r}_0^*= \bm{r}_0$ itself.

Such an initial choice of $\bm{r}_0^*$ vector is necessary, as the method is 
based on the construction of bi-orthogonal residual vectors 
$\bm{r}_{\rm i}$ and  $\bm{r}_{\rm i}^*$ recursively, for $i=1, 2, \ldots, m$,
where $m$ is the number of iterations required for convergence. The process of 
constructing the bi-orthogonal vectors gets completed, once we reach the convergence. 
In other words, the number of bi-orthogonal vectors necessary to guarantee a converged
solution represents the actual number of iterations itself. It is useful to remember that
when we refer to `bi-orthogonality' hereafter, say eg., of the residual vectors $\bm{r}_{\rm i}$,
$\bm{r}^*_{\rm i}$ we simply mean that $\la \bm{r}_{\rm i}, \bm{r}^*_{\rm j} \ra=0$ for $i \ne j$,
but  $\la \bm{r}_{\rm i}, \bm{r}^*_{\rm i} \ra$ need not be unity.\\

\noindent
(f) The bi-orthogonalization process makes use of conjugate direction vectors
$\bm{p}$ and $\bm{p}^*$ for each iteration. They can be constructed during the 
iterative process, again through recursive relations. An 
initial guess to these vectors is made as $\bm{p}_0 = \bm{r}_0$ and 
$\bm{p}_0^*= \bm{r}_0^*$.\\

\noindent
(g) The preconditioned initial residual vectors 
$\bm{\zeta}_0^*$ are computed using
\begin{equation}
\bm{\zeta}_0^*=\hat{M}^{-1} \bm{r}_0^*.
\end{equation}

\noindent
(h) For $i=1, 2, \ldots, $ the following steps are carried out 
until convergence:\\

\noindent
(i) Using the formal solver with $\bm{p}_{\rm i}$ as input (instead of actual 
source vector $\bm{y}$), $\bm {\bar J}[\bm{p}_{\rm i}]$ is obtained.\\

\noindent
(j) $\hat A[\bm{p}_{\rm i}]$ is computed using
\begin {equation}
\hat A[\bm{p}_{\rm i}]=\bm{p}_{\rm i}-(1-\epsilon)\tilde p \,\bm {\bar J}[\bm{p}_{\rm i}].
\label{A_l} \nonumber 
\end{equation}

\noindent
(k) The inner products
\begin{equation}
\la \hat{A}[\bm{p}_{\rm i}], \bm{p}_{\rm i}^* \ra \quad \textrm{and} 
\quad \la \bm{r}_{\rm i}, \bm{\zeta}_{\rm i}^* \ra ,
\label{A_l_lstar}
\end{equation}
are computed and used to estimate the quantity
\begin{equation}
\alpha_{\rm i} =\frac{\la \bm{r}_{\rm i}, \bm{\zeta}_{\rm i}^* \ra}
{\la \hat{A}[\bm{p}_{\rm i}], \bm{p}_{\rm i}^* \ra}.
\end {equation}

\noindent
(l) The new source function is obtained through
\begin{equation}
\bm{y}_{\rm i+1}=\bm{y}_{\rm i}+\alpha_{\rm i} \bm{p}_{\rm i}.
\end {equation}

{\it Test for Convergence}: Let $\bar{\omega}$ denote the convergence criteria. If 
\begin{equation}
\max_{\tau}\{\delta \bm{y}/\bm{y}\} \le \bar{\omega}, 
\end {equation}
then iteration sequence is terminated. 
Otherwise it is continued from step (m) onwards. 
The convergence criteria $\bar{\omega}$
is chosen depending on the problem.\\

\noindent
(m) Following recursive relations are used to compute the new set of
vectors to be used in the $(i+1)$th iteration:
\begin{eqnarray}
& & \bm{r}_{\rm i+1}=\bm{r}_{\rm i}-\alpha_{\rm i} \hat {A}[\bm{p}_{\rm i}], \\
& & \bm{r}_{\rm i+1}^*=\bm{r}_{\rm i}^*-\alpha_{\rm i} \hat {A}^T [\bm{p}_{\rm i}^*],\\
& & \bm{\zeta}_{\rm i+1}^*=\hat{M}^{-1} \bm{r}_{\rm i+1}^*.
\end {eqnarray}

\noindent
(n) The quantity $\beta_{\rm i}$ is computed using
\begin{equation}
\beta_{\rm i}= \frac{\la \bm{r}_{\rm i+1}, \bm{\zeta}_{\rm i+1}^* \ra}
{\la \bm{r}_{\rm i}, \bm{\zeta}_{\rm i}^* \ra}.
\end {equation}

\noindent
(o) The conjugate direction vectors for the $(i+1)$th 
iteration are computed through
\begin {eqnarray}
& &\bm{p}_{\rm i+1} = \bm{r}_{\rm i+1}+\beta_{\rm i} \bm{p}_{\rm i}, 
\nonumber \\
& &\bm{p}_{\rm i+1}^* = \bm{r}_{\rm i+1}^*+\beta_{\rm i} \bm{p}_{\rm i}^*.
\end {eqnarray}

\noindent
(p) The control is transferred to step (g). 

The converged source function $\bm{y}$ is finally used to compute the specific
intensity everywhere within the spherical medium.

\section{Transpose free variant - Pre-BiCG-STAB}
\label{prebicg-stab}
In spite of higher convergence rate, computation and storage of the 
$\hat A^T$ matrix is a main dis-advantage of the Pre-BiCG method.
To avoid this, and to make use of only the `action' of $\hat A$ matrix on an 
arbitrary vector, a method called `BiCG-squared' was developed
\citep[See][for references and details]{saad00}, which is based
on squaring the residual polynomials. Later it was improved by re-defining 
the residual polynomial as a product of two polynomials and obtaining a 
recursive relation for the new residual polynomial. This product 
involves residual polynomial of the Pre-BiCG method and a
new polynomial which `smoothens' the iterative process. In this section we
give the computing algorithm of the Pre-BiCG-STAB method as applied to 
a radiative transfer problem. As described below, we can avoid computing and
storing of the $\hat A^T$ matrix in the Pre-BiCG-STAB method. 
However we would now need to call the formal solver twice per iteration
unlike in Pre-BiCG method, where it is called only once. 
This results in an increase in number of operations per
iteration when compared to Pre-BiCG method, causing a slight increase in
the CPU time per iteration. In spite of these the Pre-BiCG-STAB method 
turns out to be always faster than the regular Pre-BiCG method in terms 
of convergence rate (lesser number of iterations for convergence).
\subsection{Pre-BiCG-STAB algorithm}
\label{stab_alg}
Now we give the algorithm of Pre-BiCG-STAB method to solve the system 
$\hat M^{-1} \hat A \bm{S} = 
\hat M^{-1} \bm{b}$.
Here $\hat M$ is a suitably chosen preconditioner matrix.
The computing algorithm is organized as follows:\\

\noindent
(a) First initial preconditioned residual vectors and conjugate direction 
vectors are defined through
\begin {equation}
\bm{z}_0=\hat M^{-1} \bm{b}-\hat M^{-1} \hat A \bm{S},
\end{equation}

\noindent
\begin {eqnarray}
\bm{z}_0^*=\bm{z}_0, \quad \bm{P}_0=\bm{z}_0.
\end{eqnarray}

\noindent
(b) For $j=1, 2, \ldots$ the following steps are carried out until 
convergence.\\

\noindent
(c) Using $\bm{P}_{\rm j}$ instead of the source function a call to the 
formal solver is made to compute $\hat A \bm{P}_{\rm j}$. \\

\noindent
(d) The coefficient $\alpha_{\rm j}$ can be evaluated now as
\begin {equation}
\alpha_{\rm j} = \frac {\la \bm {z}_{\rm j}, \bm{z}_0^* \ra}
{\la \hat M^{-1} \hat A \bm{P}_{\rm j}, \bm{z}_0^* \ra}.  
\label{prealphaj} 
\end{equation}

\noindent
(e) Another vector $\bm{q}_{\rm j}$ is calculated as
\begin{equation}
\bm{q}_{\rm j}=\bm{z}_{\rm j}-\alpha_{\rm j} \hat M^{-1} \hat A \bm{P}_{\rm j}.
\end{equation}

\noindent
(f) Using $\bm{q}_{\rm j}$ in place of the source function a call to the
formal solver is made to obtain $\hat A \bm{q}_{\rm j}$.\\

\noindent
(g) The coefficient $\omega_{\rm j}$ is estimated as
\begin {equation}
\omega_{\rm j} = \frac {\la \hat M^{-1} \hat{A}  \bm {q}_{\rm j}, \bm{q}_{\rm j} \ra}
{\la \hat M^{-1} \hat A \bm{q}_{\rm j}, \hat M^{-1} 
\hat A \bm{q}_{\rm j} \ra}.
\label{preomegaj}
\end{equation}

\noindent
(h) The updated new source function is calculated as 
\begin {equation}
\bm{S}_{\rm j+1}=\bm{S}_{\rm j}+\alpha_{\rm j} \bm{P}_{\rm j}
+\omega_{\rm j} \bm{q}_{\rm j}.
\end {equation}

\noindent
(i) Test for convergence is made as in the Pre-BiCG algorithm. \\

\noindent
(j) Before going to the next iteration a set of recursive relations are 
used to compute residual vectors 
\begin {equation}
\bm{z}_{\rm j+1}=\bm{q}_{\rm j}-\omega_{\rm j}  
\hat M^{-1} \hat A \bm{q}_{\rm j},
\end {equation}

\noindent
and conjugate direction vectors 

\begin {equation}
\bm{P}_{\rm j+1}=\bm{z}_{\rm j+1}+\beta_{\rm j}(\bm{P}_{\rm j}-\omega_{\rm j}
\hat M^{-1} \hat A \bm{P}_{\rm j}).
\end {equation}

\noindent
for the next iteration, where the coefficient $\beta_{\rm j}$ is

\begin {equation}
\beta_{\rm j} = \frac {\la \bm {z}_{\rm j+1}, \bm{z}_0^* \ra}
{\la \bm{z}_{\rm j}, \bm{z}_0^* \ra} \frac{\alpha_{\rm j}}{\omega_{\rm j}},
\label{prebetaj} 
\end{equation}

\noindent
(k) The control is now transferred to the step (b).  

\section{Comparison of ALI and Pre-BiCG methods}
\label{bicg_ali}
There are two characteristic quantities that define iterative techniques.
They are (a) convergence rate, which is nothing but the maximum relative
change (MRC) defined as
\begin {equation}
R_{\rm c}=\max_{\tau}\{\frac{\delta \bm{S}^n}{\bm{S}^n}\},
\end{equation}
and  (b) the total CPU time $T_{\rm total}$ required for convergence. 
$T_{\rm total}$ is the time taken to reach a given level of convergence,
taking account only of the arithmetic manipulations within the iteration cycle.
We also define a quantity called the true error $T_{\rm e}$ and use it
to evaluate these methods.
\subsection{The behaviour of the maximum relative change (MRC)}
In this section we compare $R_{\rm c}$ and $T_{\rm total}$ 
for the Jacobi, GS, SOR, Pre-BiCG and the Pre-BiCG-STAB methods. 
The SOR parameter used is 1.5. It is worth noting that the 
overrates (the time taken to prepare the necessary set up, before initiating the iterative cycle) are expected to be different
for different methods. For instance, in Jacobi and GS/SOR this is essentially
the CPU time required to set up the $\hat \Lambda^*$ matrix. 
In the Pre-BiCG method this involves the time taken to construct the 
$\hat A^T$ matrix, which is a critical quantity of this method. 
The Pre-BiCG method is described in this paper in the
context of a 2-level atom model, because of which, we do not need to update
the  $\hat A^T$ matrix at each iteration. For the Pre-BiCG-STAB method
it is the time taken to construct the preconditioner matrix $\hat M$.

Fig.~\ref{fig_mrc} shows a plot of $R_{\rm c}$ for different methods. 
We can take $R_{\rm c}$ as a measure of the convergence rate. \citet{che03} 
show that it always becomes necessary to use high resolution grids, 
to achieve high accuracy of the solution 
(See also Sect.~\ref{intro} of this paper). This is especially true
in the case of spherical radiative transfer where a spatial grid
with a large number of points per decade becomes necessary to achieve
reasonable accuracy. In the following we discuss how 
different methods respond to
the grid refinement. It is a well known fact with the ALI methods, that the 
convergence rate is small when the resolution of the depth grid is very high.
In contrast they have a high convergence rate in low resolution grids. On the 
other hand the $R_{\rm c}$ of Pre-BiCG and Pre-BiCG-STAB methods have
higher convergence rate even in a high resolution grid.
Fig.~\ref{fig_mrc}(a) shows $R_{\rm c}$ for different methods when a 
low resolution spatial grid is used (5pts/D in the logarithmic scale 
for $\tau$ grid). The Jacobi method has a low convergence rate. 
In comparison, GS has a convergence rate which is twice that of Jacobi. SOR 
has a rate that is even better than that of GS. 
However Pre-BiCG and the Pre-BiCG-STAB methods have the higher 
convergence rate. Fig.~\ref{fig_mrc}(b) and \ref{fig_mrc}(c) are 
shown for intermediate (8 pts/D) and high (30 pts/D) grid resolutions. 
The essential point to note is that, 
as the grid resolution increases, 
the convergence rate decreases drastically and monotonically for the Jacobi and the GS methods. 
It is not so drastic for the SOR method which shows non-monotonic dependence
on grid resolution. The Pre-BiCG and Pre-BiCG-STAB methods exhibit again a 
monotonic behaviour apart from being relatively less sensitive to 
the grid resolution. 

In Table \ref{table_1} we show what happens when we set convergence 
criteria to progressively smaller values 
($\bar \omega=$$10^{-6}$, $10^{-8}$, and $10^{-10}$ for Tables 
\ref{table_1a}, \ref{table_1b} and \ref{table_1c}
respectively) for various grid resolutions. 
The model used to compute these results is
($\tilde n$, $R$, $T$, $a$, $\epsilon$, $\beta_{\rm c}$, $B_{\nu}$)=
(0, 10, $10^{3}$, $10^{-3}$, $10^{-4}$, 0, 1).
The idea is to demonstrate that for a given grid 
resolution (corresponding rows of the Tables \ref{table_1a}, 
\ref{table_1b} and \ref{table_1c}),  all the methods show 
a monotonic increase in the number of iterations for convergence,
as we decrease the $\bar \omega$.  On the other hand Pre-BiCG 
and Pre-BiCG-STAB require much less number of iterations to reach
the same level of accuracy.

{\it CPU time considerations}: Table~\ref{table_2} shows the 
CPU time requirements for the methods discussed 
in this paper. The model used to compute these test cases is 
($\tilde n$, $R$, $T$, $\epsilon$, $\beta_{\rm c}$, 
$B_{\rm \nu}$, $\bar \omega$)=
(0, 300, $10^3$, $10^{-4}$, 0, 1, $10^{-8}$). The grid 
resolution considered is 30 pts/D. The CPU time for convergence
can be defined as the computing time required to complete the
convergence cycle and reach a fixed level of accuracy. We
recall that the overrates in computing time is the time taken
to prepare the necessary set up before initiating the iterative cycle.
Total computing time is the sum of these two. In appendix 
\ref{appendix_lambda}
we discuss in detail how to construct $\hat \Lambda^*$ matrix for Jacobi, 
GS/SOR methods and $\hat A$ and $\hat M$ matrices for Pre-BiCG and 
Pre-BiCG-STAB methods respectively with an optimum effort. 
Construction of these matrices constitutes the overrates in computing time 
of each method. The first row of Table~\ref{table_2} shows that 
Pre-BiCG is the fastest to complete the convergence cycle. The reason why
Pre-BiCG-STAB takes slightly longer time than Pre-BiCG is explained
at the end of Sect.~\ref{prebicg-stab}.

The second row of Table~\ref{table_2} shows that all methods except Pre-BiCG
take nearly 8 seconds as overrates for the chosen model. Pre-BiCG
takes additional 3-4 seconds as explicit integrals are performed
for computing off-diagonal elements also (Unlike the other methods
where such integrals are performed only for diagonal elements).

The last row of Table~\ref{table_2} shows that in terms of total
CPU time requirement, the other methods fall behind the Pre-BiCG
and the Pre-BiCG-STAB. Pre-BiCG seems to be a bit faster compared
to Pre-BiCG-STAB for the particular model chosen. However it is model
dependent. For instance, as the contribution towards overrates increases,
Pre-BiCG-STAB clearly stands out as the fastest method of all, discussed
in this paper.

\subsection{A study of the True Error}
We now study the true errors in these methods (see Figure \ref{fig_true}).
The model parameters are
($\tilde n$, $R$, $T$, $\epsilon$, $\beta_{\rm c}$, $B_{\nu}$)=
(0, 10, $10^{3}$, $10^{-4}$, 0, 1). A coherent scattering
limit is used.
To define a true error, we need a so called `exact solution'. Except for
highly idealized cases, exact solutions do not exist. For practical
purposes, the exact solution can be defined as a solution obtained on a
spatial grid of resolution that is three times larger than the grid 
resolution of the model that we are interested in. Also, we extend the 
iteration until $R_{\rm c}$ reaches an extremely small value of $10^{-12}$.
The source function computed in this way can be called $\bm{S}(\infty,\infty)$
(fully converged solution on an infinite resolution) \citep[see][]{aue94}.
The source function at the $n$th iterate is denoted by $\bm {S}^{n}$.
We define the true error as
\begin {equation}
T_{\rm e} = \max_{\tau} \mid \frac{\bm {S}^{n}-\bm{S}_{\rm exact}}
{\bm{S}_{\rm exact}} \mid,
\end{equation}
following \citet{tru95}. 
In Fig. \ref{fig_true}(a) we show $T_{\rm e}$
computed for the Pre-BiCG method using three grid resolutions,
namely 10 pts/D, 14 pts/D and 20 pts/D. 
The plateau of each curve represents
the minimum value of the true error reached for a given grid resolution. 
We notice that as the resolution increases, $T_{\rm e}$ 
gradually decreases in magnitude as expected. 
In Fig.~\ref{fig_true}(b) we show $T_{\rm e}$ computed for the Pre-BiCG-STAB
method. The model parameters are same as in Fig.~\ref{fig_true}(a).
Clearly, Pre-BiCG-STAB shows a smooth decrement of true error 
compared to Pre-BiCG, because of the smoothing polynomial used to 
define the residual vectors.
In Fig.~\ref{fig_true}(c) we compare the decrement of true errors for different
iterative methods.
The grid resolution chosen is 14 pts/ D with other model parameters
being same as in Figs.~\ref{fig_true}(a), (b). The decrease of the true errors
follows the same pattern in all the iterative methods, although the number of
iterations required for $T_{\rm e}$ to reach a constant value (plateau)
depends on the method. To reach the same level of true error, the Pre-BiCG
and Pre-BiCG-STAB methods require considerably less number of iterations, 
when compared to the other three. 

\subsection{ A theoretical upper bound on the number of iterations for convergence
in the Pre-BiCG method}
Suppose that $\hat A$ is an $N_{\rm d} \times N_{\rm d}$ matrix. 
The solution to the problem $\hat A \bm{y}=\bm{b}$ is a vector of length $N_{\rm d}$. 
In an $N_{\rm d}$-dimensional vector space $V_{\rm N_{\rm d}}$
the maximum number of linearly independent vectors is $N_{\rm d}$. Hence, there
can at the most be $N_{\rm d}$ orthogonal vectors in $V_{\rm N_{\rm d}}$. 
The Pre-BiCG method seeks a solution by constructing orthogonal vectors. 
We recall that the residual counterpart vectors $\{ \bm{r}_1^*, \bm{r}_2^*, 
\ldots, \bm{r}_M^*\}$ constructed during the iteration process are 
orthogonal to the initial residual vector $ \bm{r}_0$. Thus, when we 
reach convergence after $M$ iterations, we will have a set of $M+1$
orthogonal vectors $\{\bm{r}_0, \bm{r}_1^*, \bm{r}_2^*, \ldots, 
\bm{r}_M^*\}$. From the arguments given above, it is clear that $M+1 \le 
N_{\rm d}$, namely in the Pre-BiCG method, `the convergence must be 
reached theoretically in at the most $N_{\rm d}$ steps (or iterations)'. This
sets an upper limit to the number of iterations to reach convergence 
\citep[see also][]{hes52}. For example when the dimensionality of a problem 
is high (very large value of $N_{\rm d}$), the Pre-BiCG method ensures 
convergence in at the most $N_{\rm d}$ iterations. A theoretical upper
bound on the number of iterations also exists for the Pre-BiCG-STAB method, 
whereas the other methods do not have such a theoretical upper 
bound. In practice we find that Pre-BiCG and Pre-BiCG-STAB methods 
actually require much less number of iterations than $N_{\rm d}$, 
even when $N_{\rm d}$ is large. 
\section{Results and discussions}
\label{results}
The main purpose of this paper is to propose a new method to solve the line
transfer problems in spherically symmetric media. In this section we show some
illustrative examples in order to compare with the famous benchmarks for spherical
transfer solutions presented in the papers by \citet{kun74}.
In Fig.~\ref{fig_source} we show source functions for different test cases.
Fig.~ \ref{fig_source}(a) shows the source functions for $\epsilon=10^{-2}$ and 
$\epsilon=10^{-4}$ for $R=300$ and $T=10^3$. 
Other model parameters are ($\tilde n$, $\beta_{\rm c}$, $B_{\rm \nu}$)=(2, 0, 1). 
We use a Doppler profile
to compare with the results of \citet{kun74}. Plane parallel result 
is also shown for comparison. When $\epsilon=10^{-2}$ we observe that the 
thermalization is reached at the thermalization length for the Doppler profile
namely $1/\epsilon=100$. When $\epsilon=10^{-4}$ thermalization 
does not occur. Clearly the minimum value of the source function 
is $\epsilon B_{\nu}$. For the large values of $R=300$ 
and opacity index $\tilde{n}=2$, as $r$ increases the 
opacity decreases steadily and the source function indeed
approaches this minimum value near the surface layers. 
For this case, the departure of the source function from planar 
limit is severe near the surface. It can be shown 
(dashed line of Fig.~\ref{fig_source}(b), see also Fig.~3 of \citet{kun74}) that
this departure is not so acute when $\tilde{n}=0$, but is more acute when $\tilde{n}=3$
(dash triple-dotted line in Fig.~\ref{fig_source}(b)). 
In  Fig.~\ref{fig_source}(b) we plot source
function for the same model as Fig.~\ref{fig_source}(a) but for various values of
$\tilde{n}$. For negative $\tilde{n}$, the distinction between $\bm{S}(\tau)$ vs. $\tau$
curves for different $\tilde n$ is small. For positive $\tilde{n}$, the effects
are relatively larger (see dot-dashed and long dashed curves in Fig.~\ref{fig_source}(b)).

In Fig.~\ref{fig_source}(c), we show
source function variation for a range of spherical extensions $R$. We have chosen 
an effectively optically thin model ($T$, $\epsilon$)=($10^{8}$, $10^{-10}$) 
because in such a medium, thermalization effects do not completely
dominate over the effects of sphericity. Other parameters are same as in 
Fig.~\ref{fig_source}(b). 
Clearly, the decrease in the value of source function throughout the atmosphere
is monotonic, with an increase in the value of $R$ from 1 to $10^6$.

In Fig.~\ref{fig_intensity1}, we show effects of limb darkening in spherical 
atmospheres for $R=10^3$ and $R=10^6$. The other model parameters 
for Fig.~\ref{fig_intensity1}(a) are 
($\tilde n$, $T$, $\epsilon$, $\beta_{\rm c}$, $B_{\nu}$)=
(2, $10^8$, $10^{-4}$, 0, 1). A Doppler line profile is used.
From the Figure, we notice absorption in 
the line core and emission in the near line wings ($x \approx 4$) 
for $\theta=0^\circ$ and $10^\circ$.
This is the characteristic self reversal observed in spectral lines formed in 
extended spherical atmospheres. The self reversal 
decreases gradually as $\theta$ increases, and finally
vanishes for large values of $\theta$. Indeed for extreme value of 
$\theta=90^{\circ}$, we observe a pure emission line. 

In Fig.~\ref{fig_intensity1}(b) we show line profiles formed in a semi-infinite 
spherical medium. The model parameters are same as in Fig.~\ref{fig_intensity1}(a)
except for ($R$, $T$)=(300, $10^{12}$). The profiles for a range of
$\theta=0^{\circ}, 31^{\circ}, 54^{\circ}, 72^{\circ}, 84^{\circ}$
are shown. For the core rays ($\theta=0^{\circ}$) we see a pure absorption line due to
thermalization of source function. For other angles,
as expected we see chromospheric type self-reversed emission lines, formed in
the lobe part of the spherical medium.
\section{Conclusions}
\label{conclusions}
In this paper we propose a robust method called Preconditioned Bi-Conjugate 
Gradient (Pre-BiCG) method to solve the classical problem of line transfer in 
spherical media. This method belongs to a class of iterative methods based on 
the projection techniques. We briefly present the method, and the computing 
algorithm. We also present a transpose-free variant called the Stabilized 
Preconditioned Bi-Conjugate Gradient (Pre-BiCG-STAB) method which is more 
advantageous in some of its features. The Pre-BiCG and Pre-BiCG-STAB methods 
are validated in terms of its efficiency and accuracy, by comparing with 
the contemporary iterative methods like Jacobi, GS and SOR.
To calculate the benchmark solutions we use spherical shell atmospheres. 
Few difficult test cases are also presented to show that the Pre-BiCG and
Pre-BiCG-STAB are efficient numerical methods for spherical line transfer.

\begin{acknowledgements}
L. S. Anusha likes to thank Dr. Han Uitenbroek, Dr. A. Asensio Ramos and Dr. M. Sampoorna for useful
discussions.
\end{acknowledgements}
%%%%%%%%%%%%%%%%%%%%%%%%%%%%%%%%%%%%%%%%%%%%%%%%%%%%%%%%%%%%%%%%%%%%%%%%%%%%%%%%%%%

%%%%%%%%%%%%%%%%%%%%%%%%%%%%%%%%%%%%%%%%%%%%%%%%%%%%%%%%%%%%%%%%%%%%%%%%%%%%%%%%%%
%%%%%%%%%%%%%%%%%%%%%%%%%%%%%%%%%%%%%%%%%%%%%%%%
\begin{figure*}
\centering
\includegraphics[scale=0.6,angle=270]{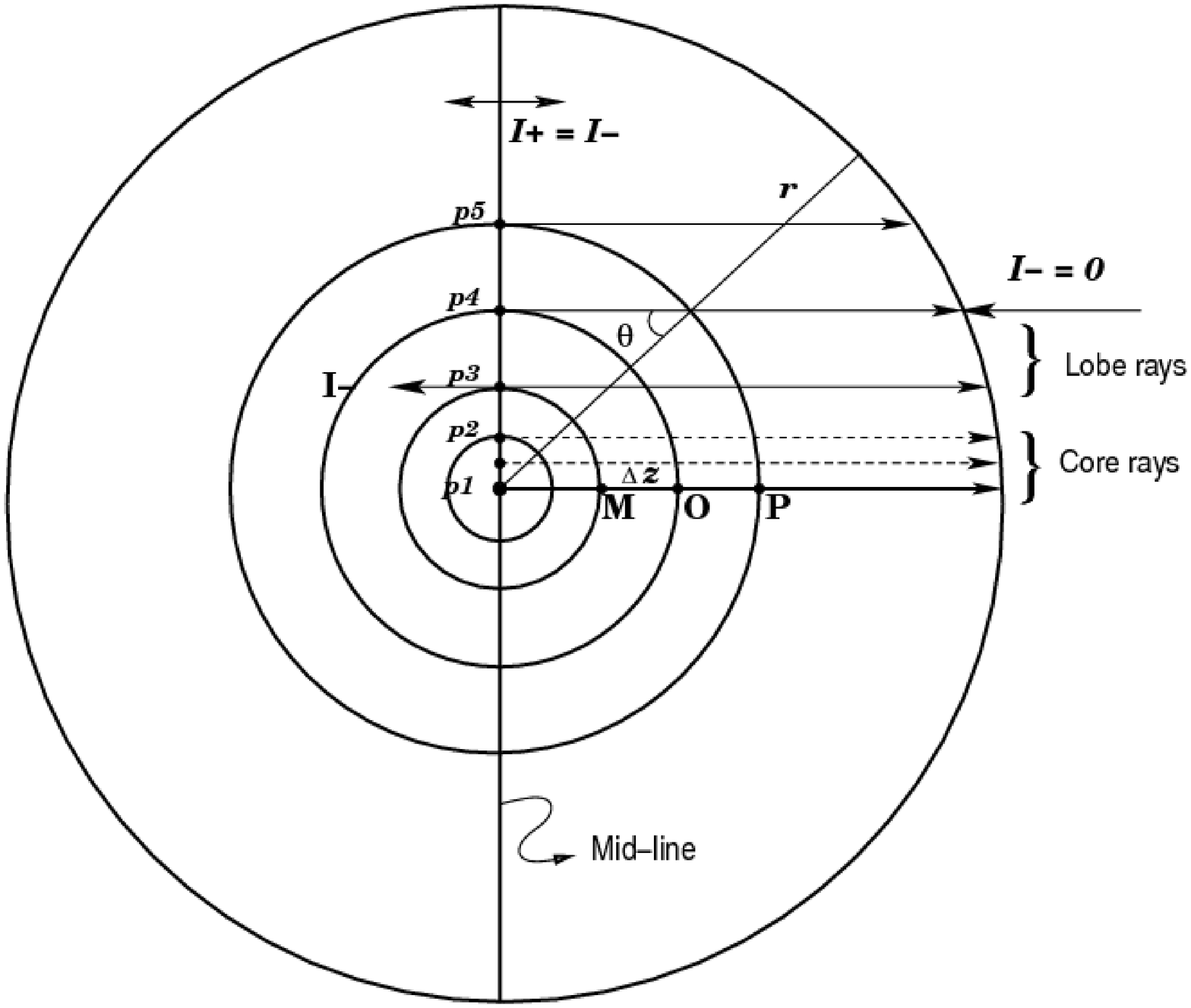}
\caption{Geometry of the problem showing the computation of radiation
field in a spherically symmetric system. The set of core 
rays and tangent rays are marked. The core is defined as a 
sphere with radius $r=1$ in units of the core radius $R_{\rm core}$.
The surface is a sphere of radius $r=R$ in units of $R_{\rm core}$.
The rays that intersect the core are called `core rays', and the rest are called
`lobe rays'. No radiation is incident on the outer 
surface of the sphere (outer boundary condition).
For all the examples presented in this paper, we use a reflecting boundary condition at the
$z=0$ vertical axis (the mid-line), namely same inner 
boundary conditions are used for both the core and the lobe rays.}
\label{fig_geometry}
\end{figure*}
%%%%%%%%%%%%%%%%%%%%%%%%%%%%%%%%%%%%%%%%%%%%%%%%
%%%%%%%%%%%%%%%%%%%%%%%%%%%%%%%%%%%%%%%%%%%%%%%%fig
\begin{figure*}
\centering
\includegraphics[scale=0.3]{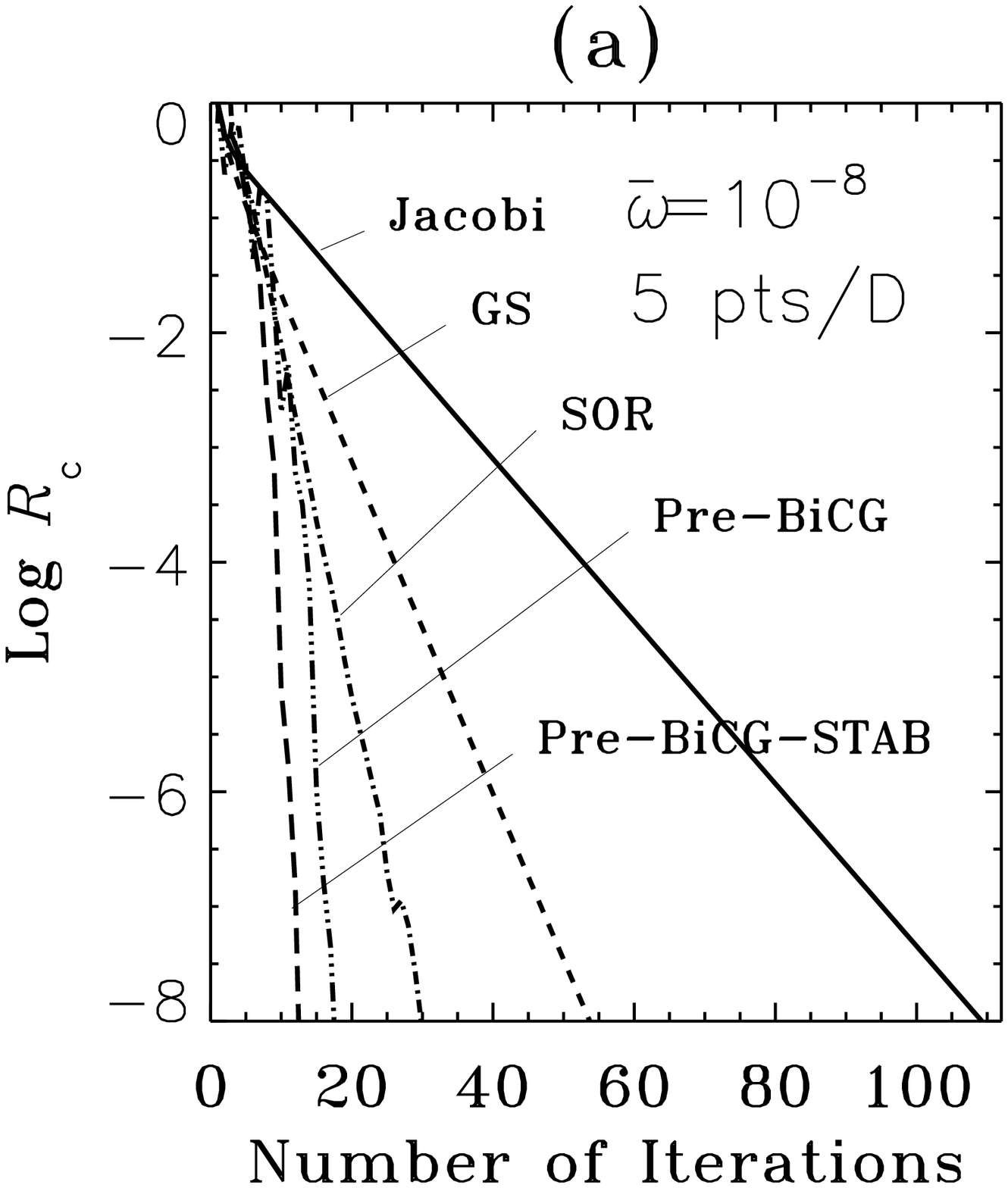}
\hspace{0.2cm}
\includegraphics[scale=0.3]{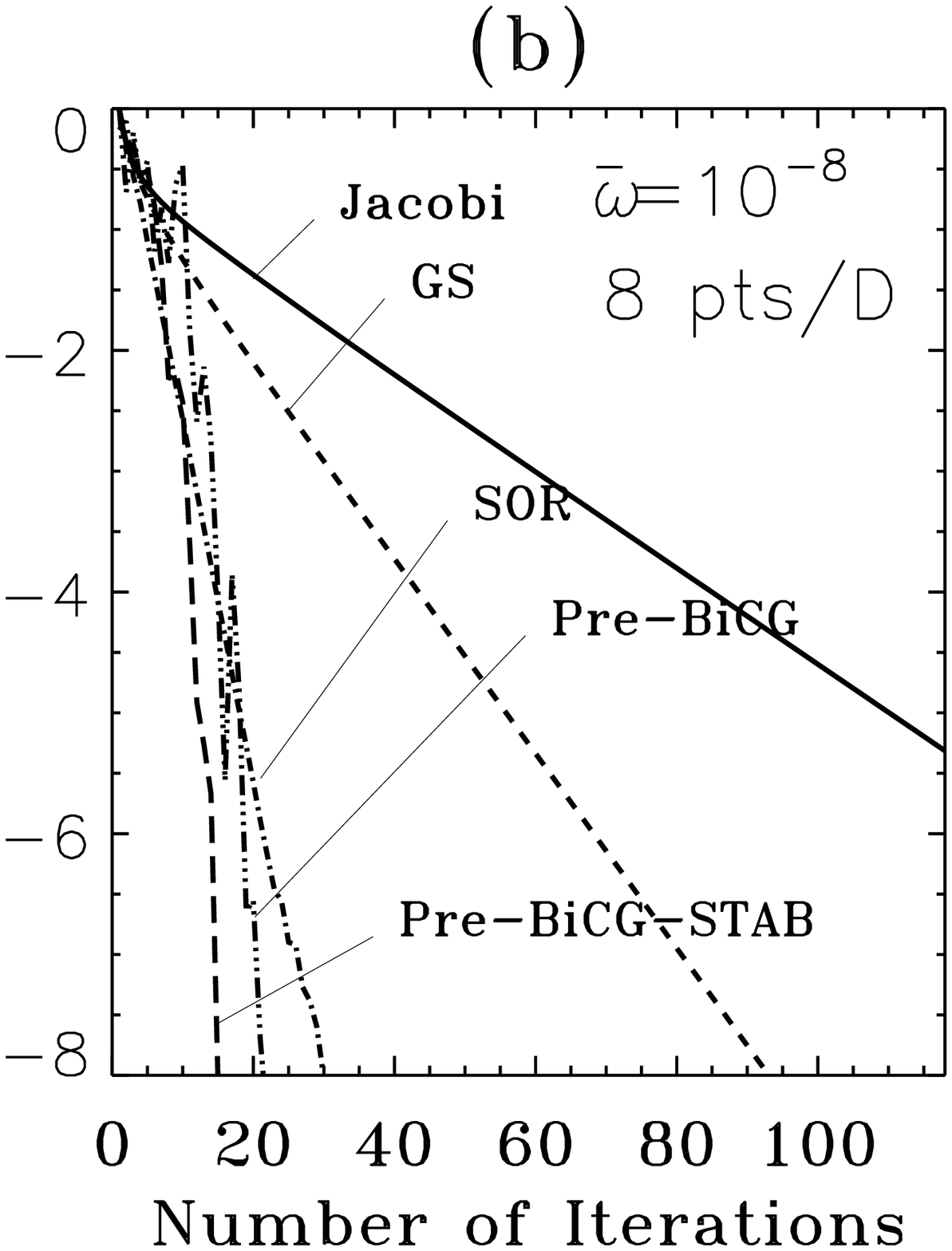}
\hspace{0.2cm}
\includegraphics[scale=0.3]{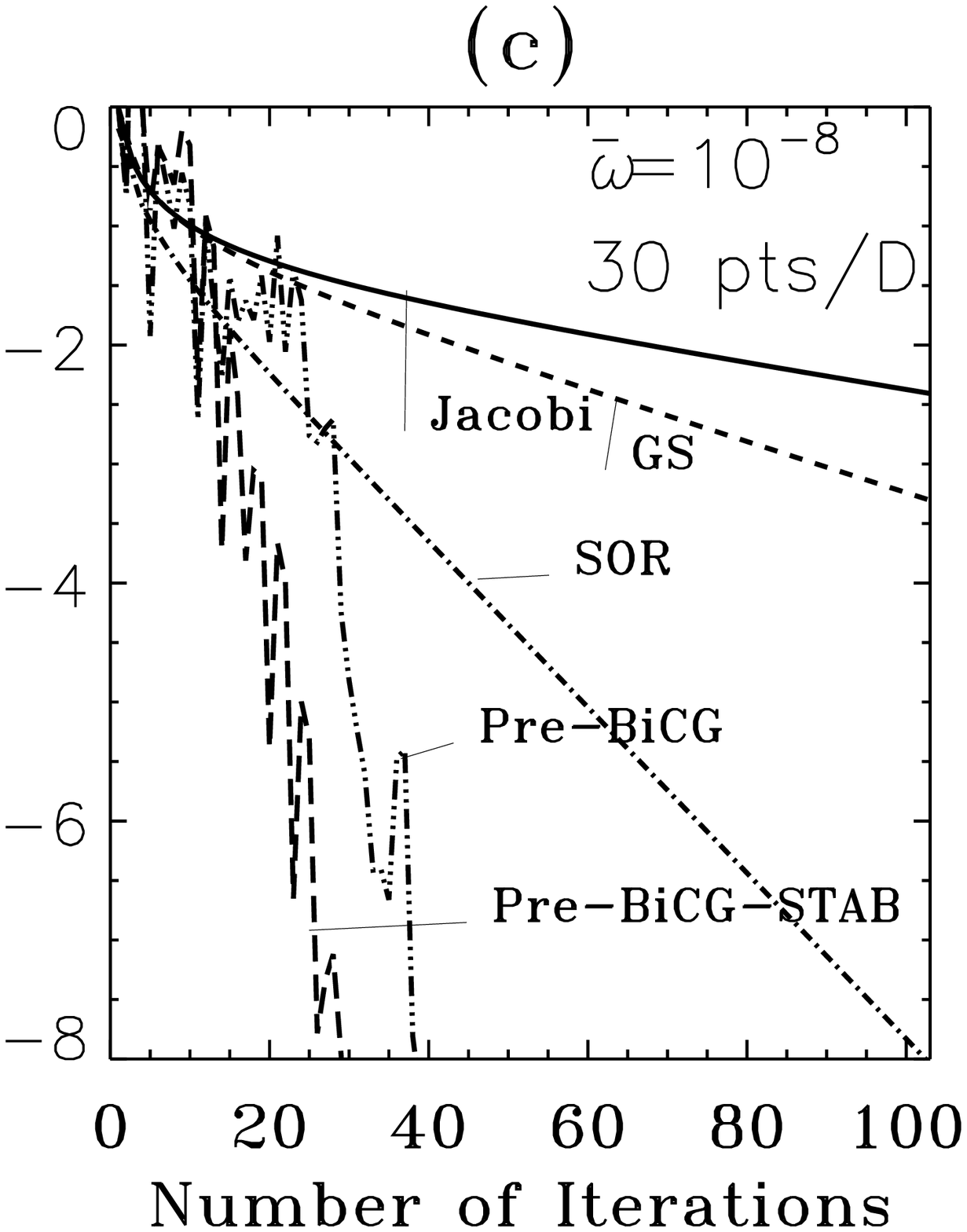}
\caption[]{Dependence of the Maximum Relative Change $R_{\rm c}$ on the iterative progress
for different methods. Panels (a), (b), and (c) represent models with low,
medium and high spatial resolution respectively.  The model parameters are
($\tilde n$, $R$, $T$, $a$, $\epsilon$, $\beta_{\rm c}$, $B_{\nu}$)=
(0, 10, $10^3$, $10^{-3}$, $10^{-4}$, 0, 1).
The convergence criteria is chosen arbitrarily as
$\bar \omega=10^{-8}$. The SOR parameter $\omega=$1.5. 
The figures show clearly that Jacobi method has the smallest convergence rate,
which progressively increases for GS and SOR methods. Pre-BiCG and Pre-BiCG-STAB
methods generally have the largest convergence rate 
compared to the other three.}
\label{fig_mrc}
\end{figure*}
%%%%%%%%%%%%%%%%%%%%%%%%%%%%%%%%%%%%%%%%%%%%%%%%fig 
%%%%%%%%%%%%%%%%%%%%%%%%%%%%%%%%%%%%%%%%%%%%%%%%fig 
\begin{figure*}
\centering
\includegraphics[scale=0.3]{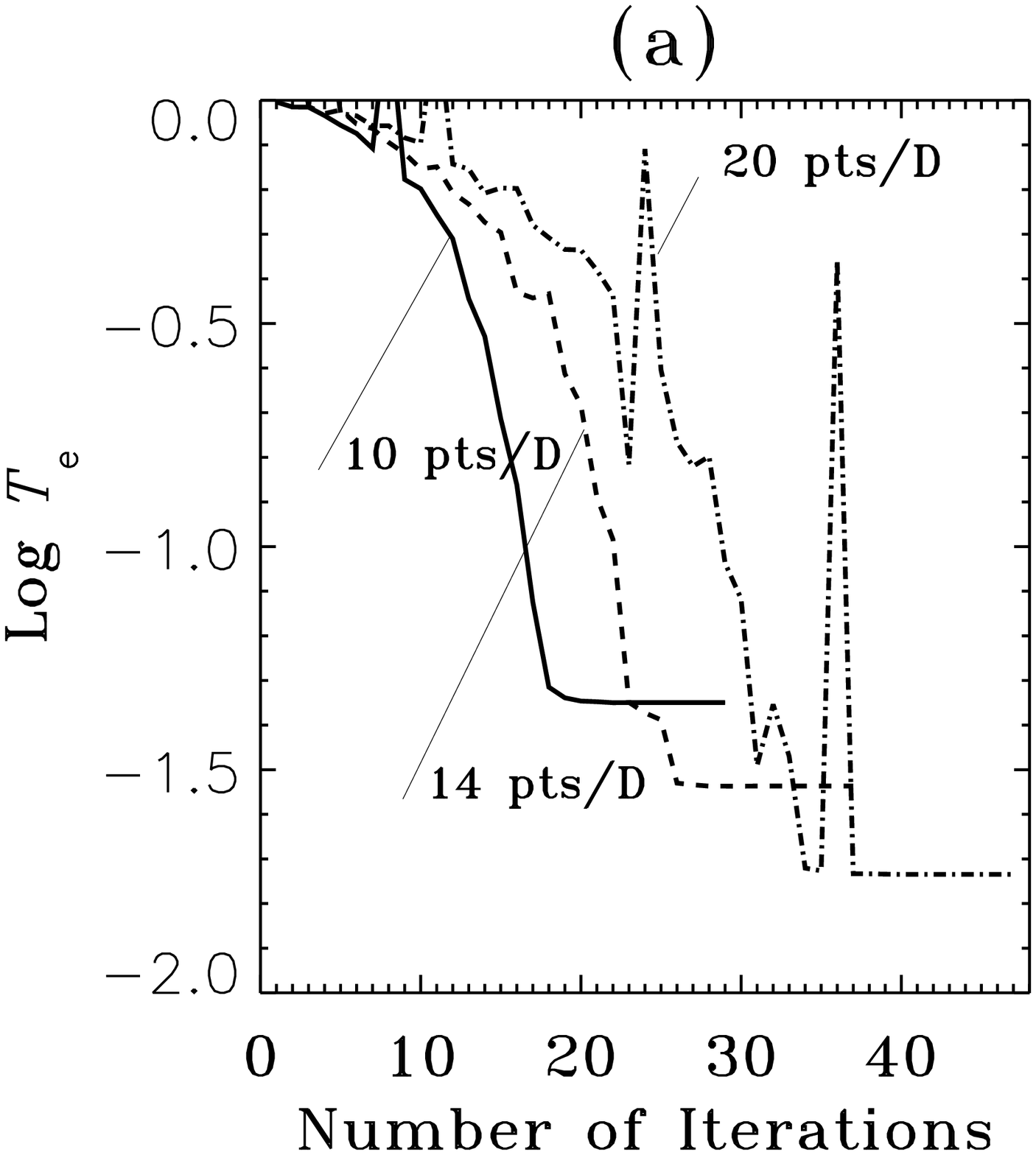}
\hspace{0.1cm}
\includegraphics[scale=0.3]{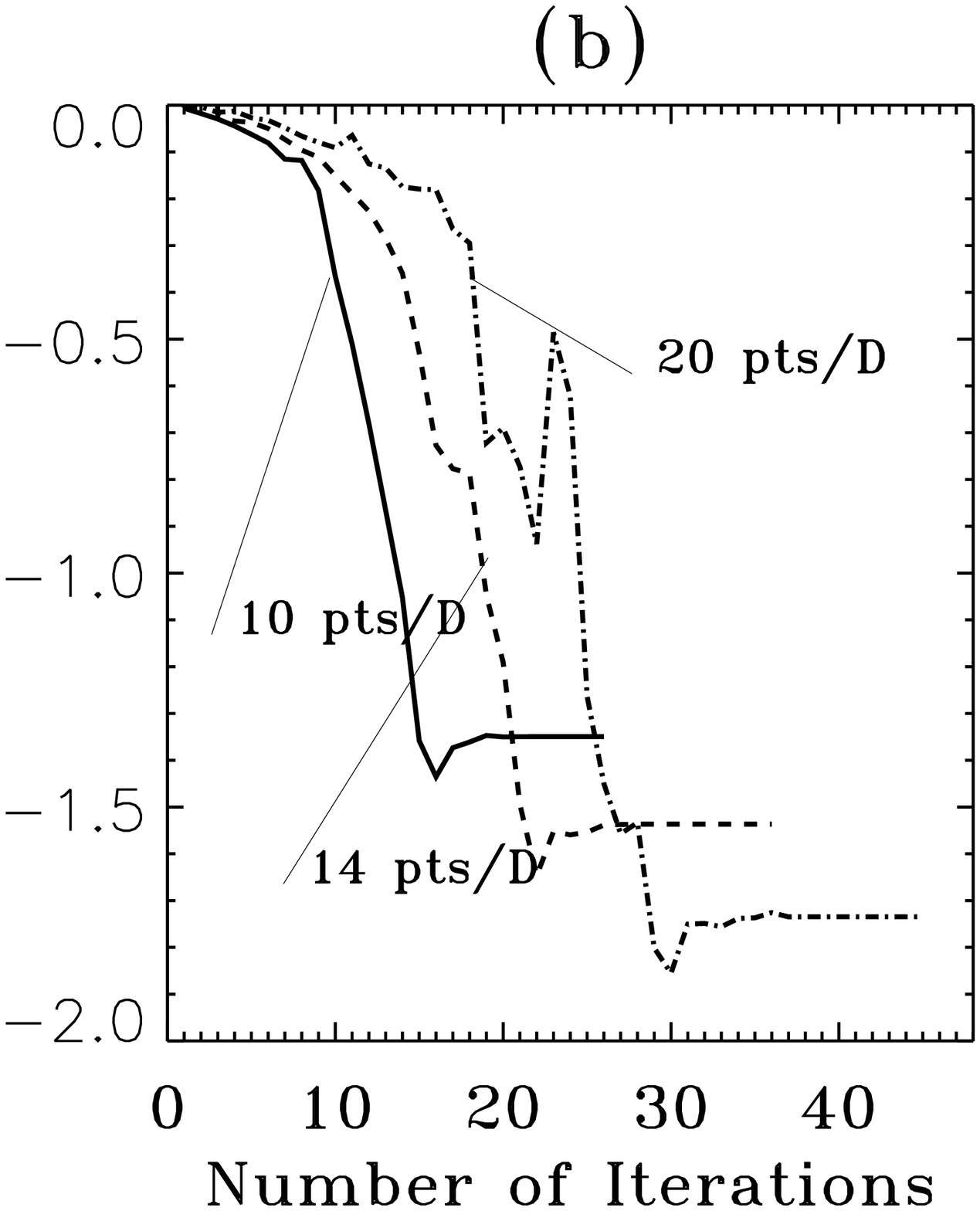}
\hspace{0.1cm}
\includegraphics[scale=0.3]{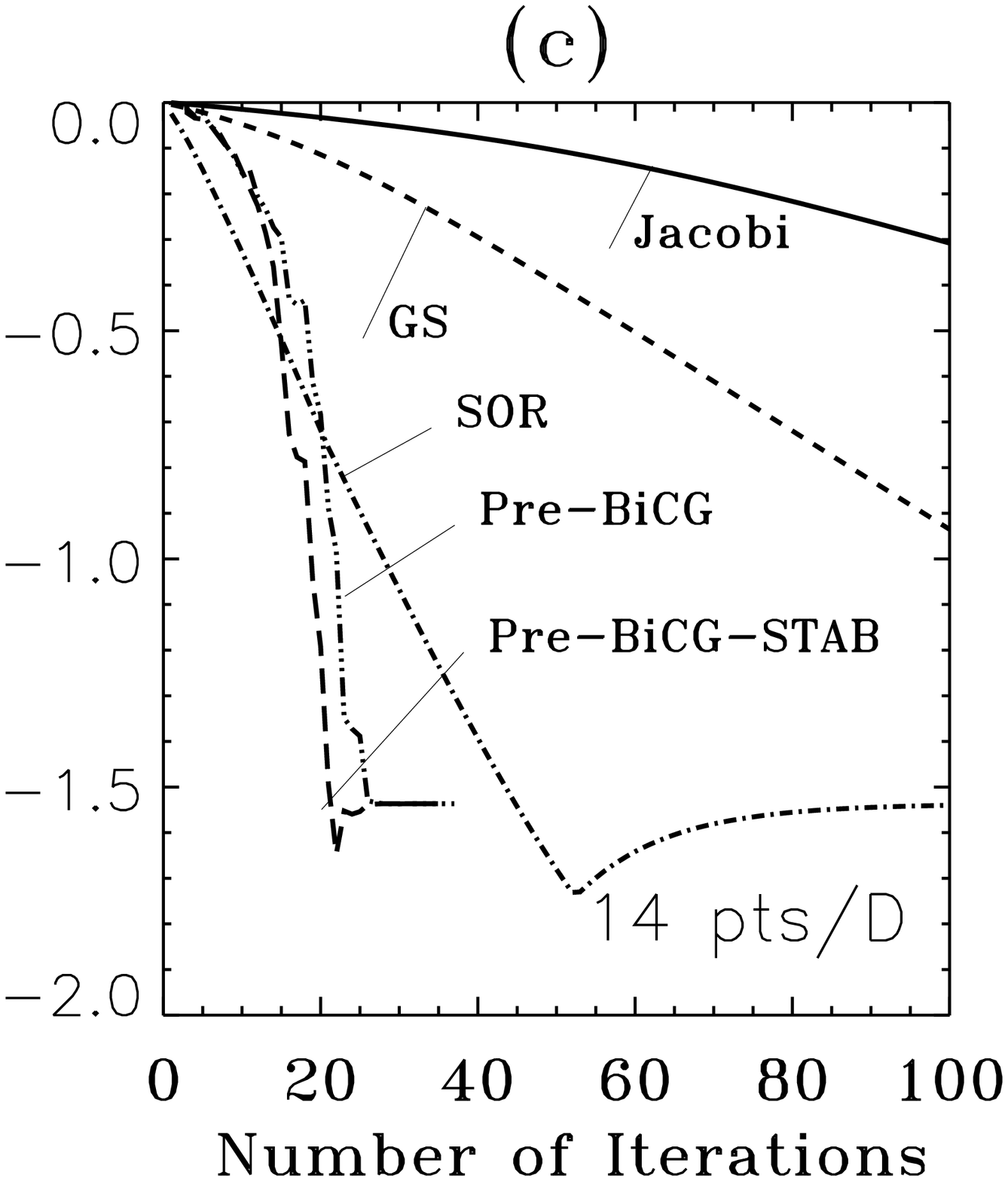}
\caption[]{Behaviour of the true error $T_{\rm e}$ in a spherically
symmetric medium. The model parameters are 
($\tilde n$, $R$, $T$, $\epsilon$, $\beta_{\rm c}$, $B_{\nu}$)=
(0, 10, $10^{3}$, $10^{-4}$, 0, 1). 
Panel (a) shows the decrement of the true error
of the Pre-BiCG method for three different spatial grid resolutions. Notice the
plateau in the true error. Panel (b) shows $T_{\rm e}$ for
Pre-BiCG-STAB method. Panel (c) shows for different methods, the 
number of iterations required to reach a constant true error 
$2.9 \times 10^{-2}$. The SOR parameter $\omega=$1.5. 
The overall behaviour of the curves is same for all the methods, although the
rates of decrement are different.}
\label{fig_true}
\end{figure*}
%%%%%%%%%%%%%%%%%%%%%%%%%%%%%%%%%%%%%%%%%%%%%%%%fig
%%%%%%%%%%%%%%%%%%%%%%%%%%%%%%%%%%%%%%%%%%%%%%%%fig 
\begin{figure*}
\centering
\includegraphics[scale=0.3]{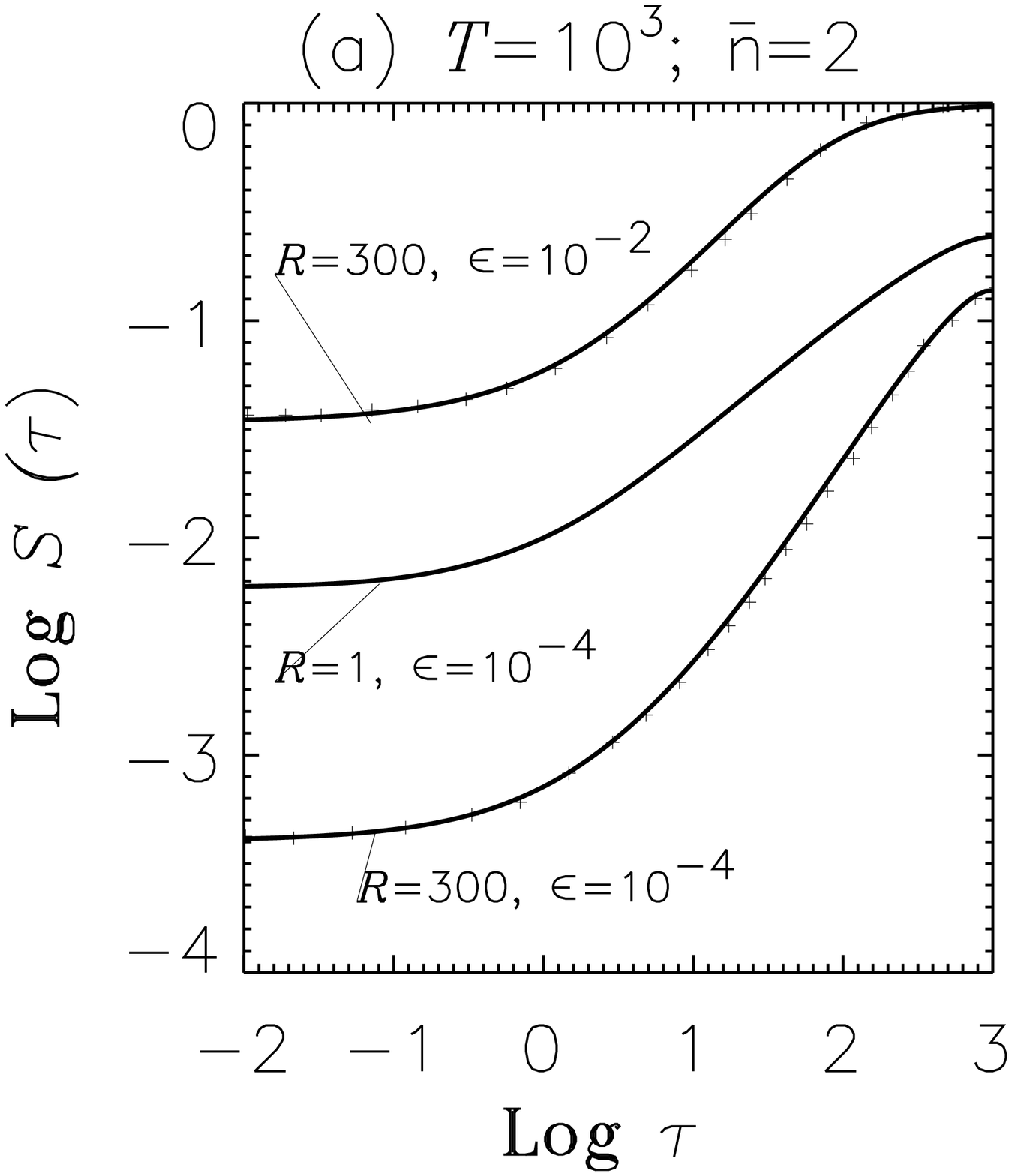}
\hspace{0.2cm}
\includegraphics[scale=0.3]{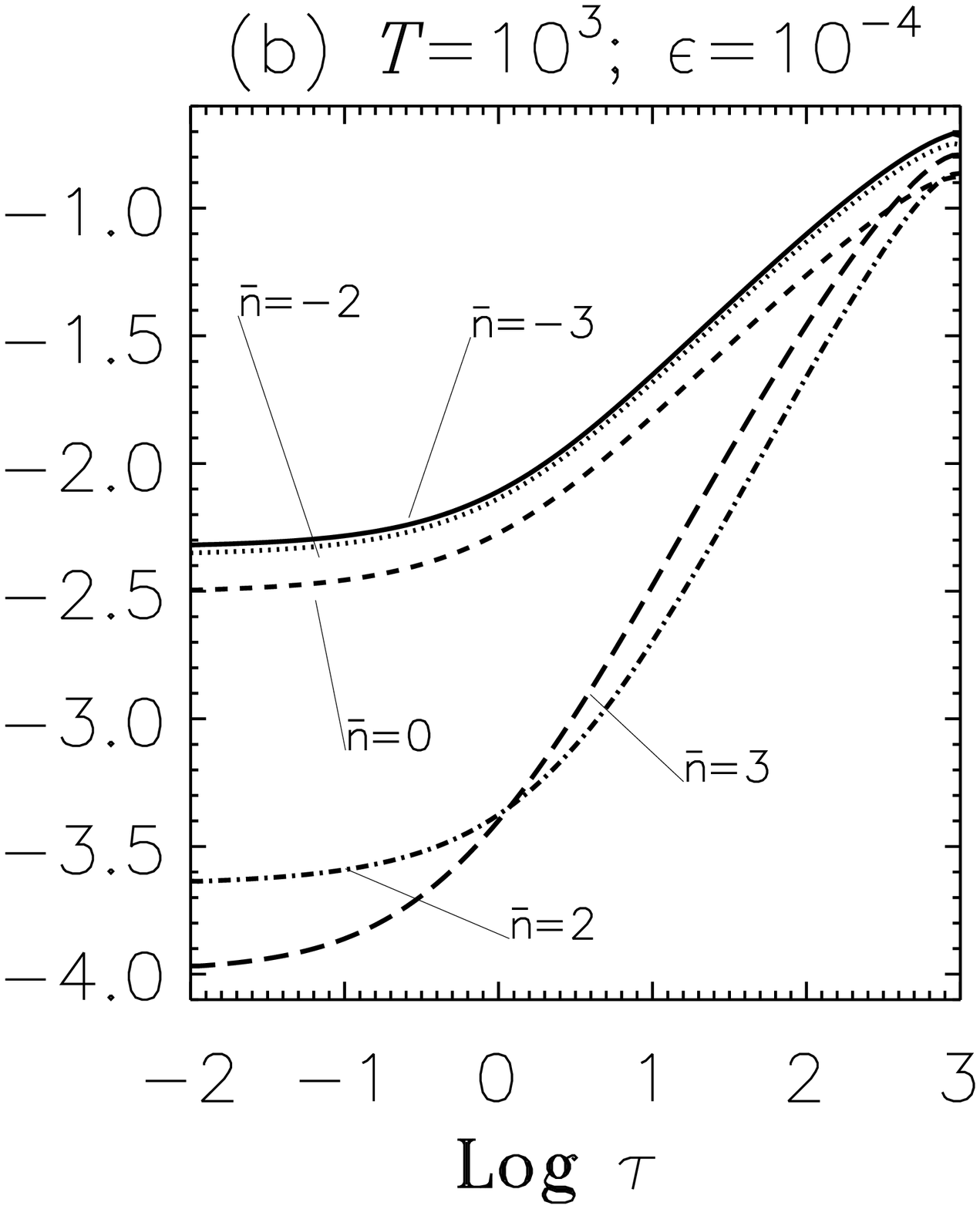}
\hspace{0.2cm}
\includegraphics[scale=0.3]{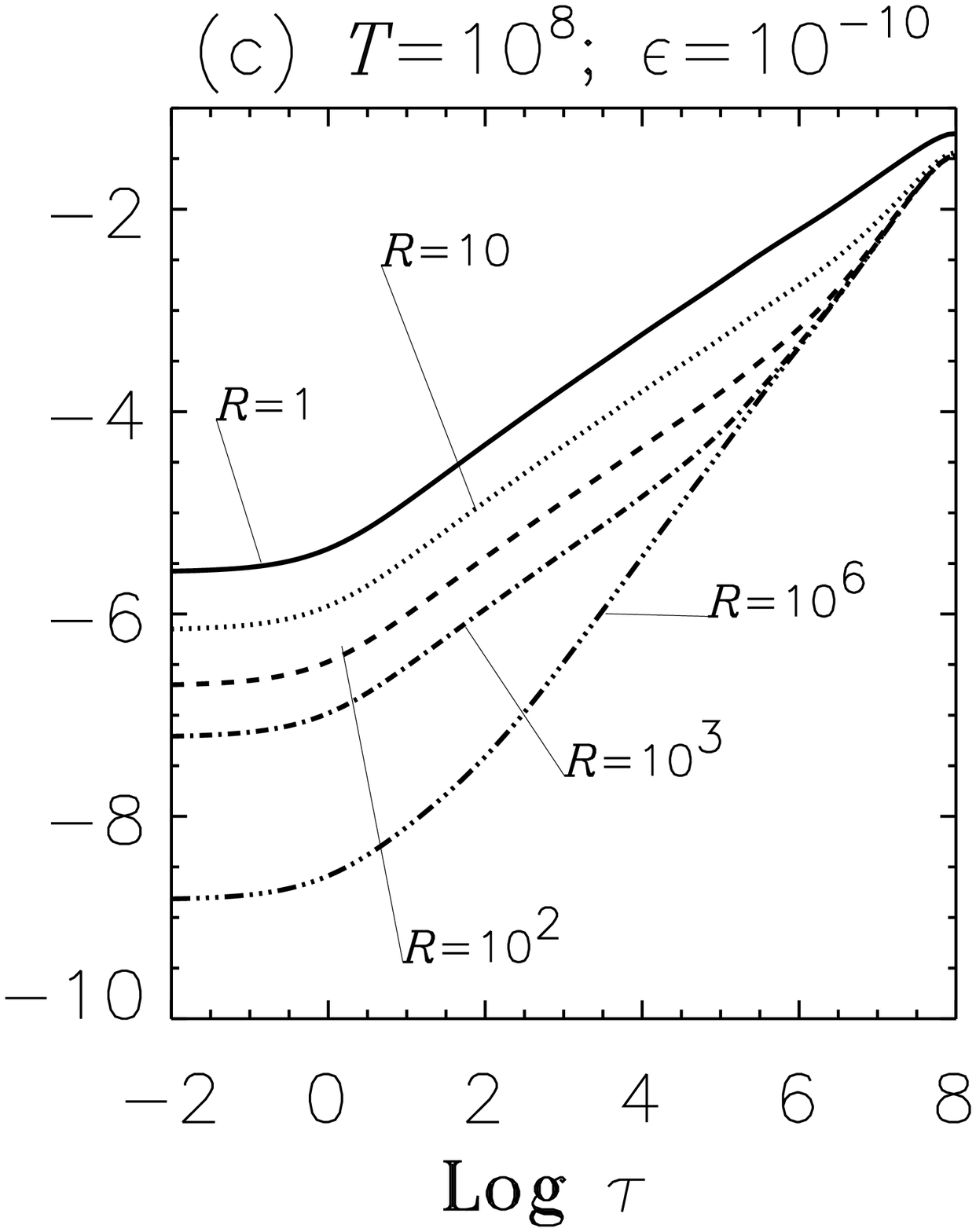}
\caption[]{In panel (a) the source function variation with optical depth is shown
for a spherical media with inverse square opacity variation for two different values of 
$\epsilon$. The symbols show the benchmark solution read from \citet{kun74},
which compare well with the solution by our method (Pre-BiCG - full lines).
The plane parallel solution ($R=1$) is shown for comparison.
Panel (b) shows the effect of power law opacity indices $\tilde n$ 
on the source function variation with $\tau$. In panel (c) the effects of 
spherical extension $R$ are shown by taking a difficult case of highly 
scattering, effectively optically thin medium.}
\label{fig_source}
\end{figure*}
%%%%%%%%%%%%%%%%%%%%%%%%%%%%%%%%%%%%%%%%%%%%%%%%fig 
%%%%%%%%%%%%%%%%%%%%%%%%%%%%%%%%%%%%%%%%%%%%%%%%fig
\begin{figure*}
\centering
\includegraphics[scale=0.38]{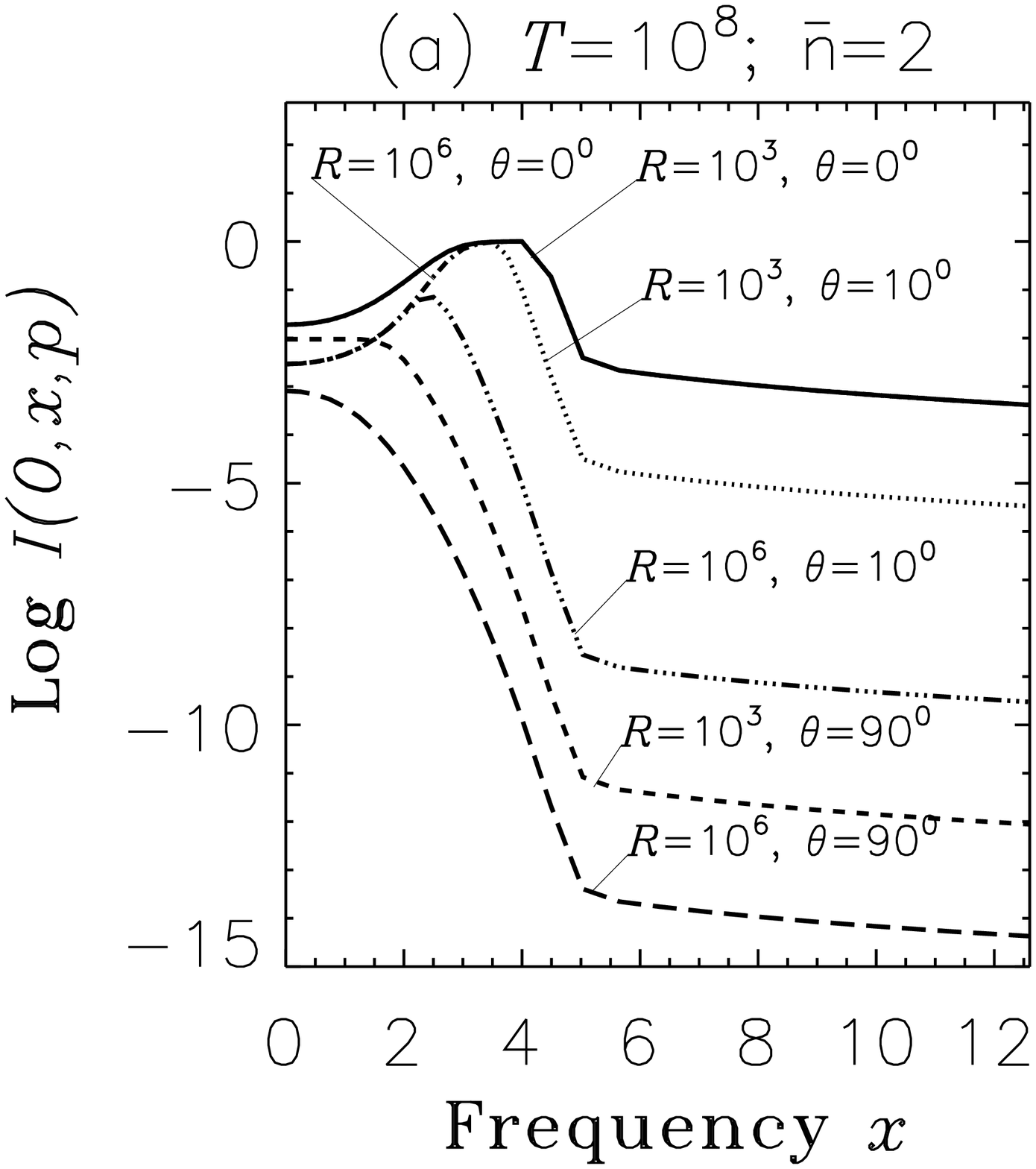}
\hspace{0.2cm}
\includegraphics[scale=0.38]{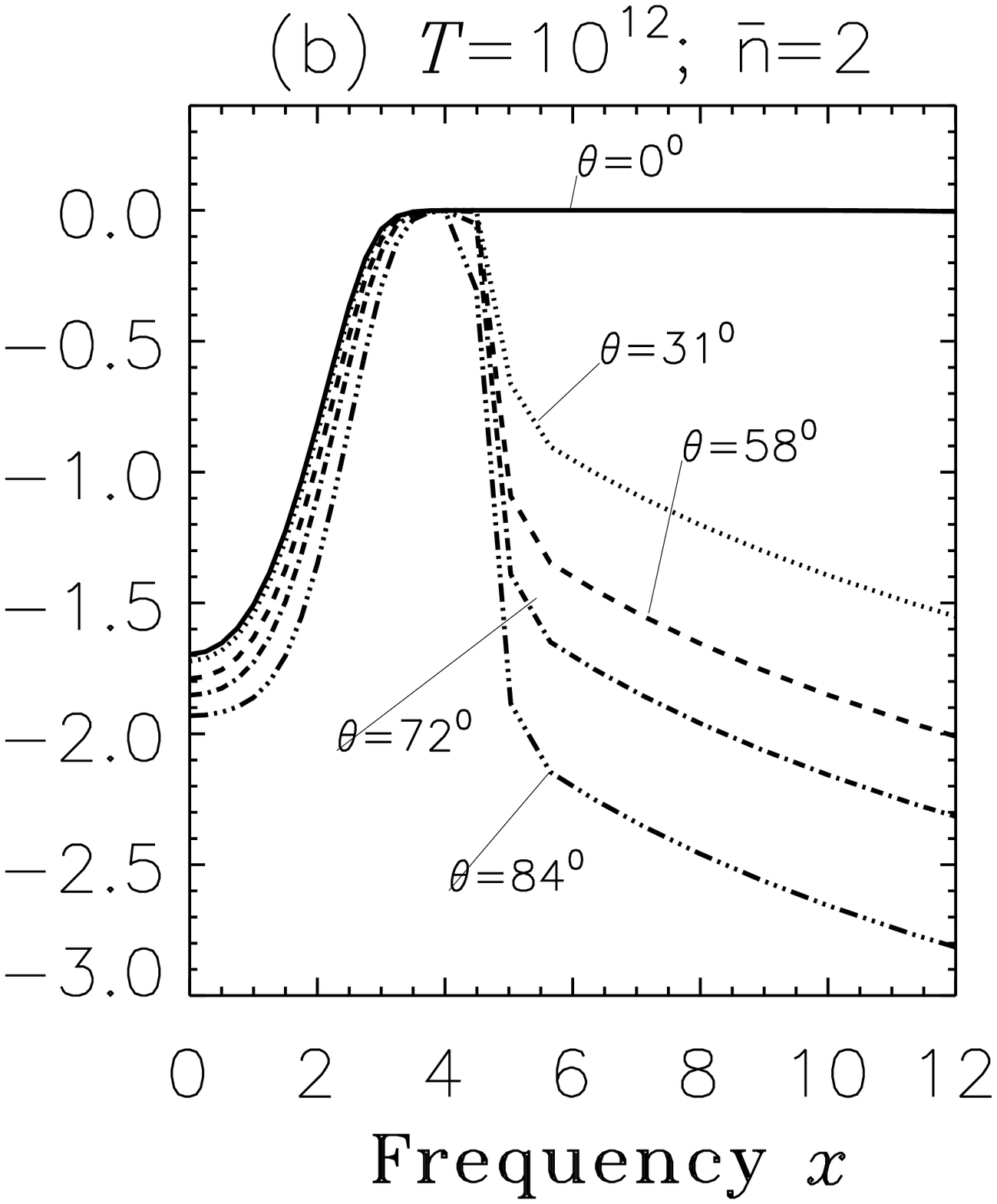}
\caption[]{Angular dependence of emergent intensities in highly extended spherical media 
($R=10^3$ and $R=10^6$). In panel (a) the profiles for the 
central ray ($\theta=0^{\circ}$), and the lobe rays ($\theta=10^{\circ}, 90^{\circ}$) are shown. 
Panel (b) shows the line profiles formed in a semi-infinite spherical 
atmosphere  with $R=300$ and $T=10^{12}$.}
\label{fig_intensity1}
\end{figure*}
%%%%%%%%%%%%%%%%%%%%%%%%%%%%%%%%%%%%%%%%%%%%%%%%%%%%%fig
%%%%%%%%%%%%%%%%%%%%%%%%%%%%%%%%%%%%%%%%%%%%%%%%%table
\begin{table*}
\begin{center}
\caption{The sensitivity of different iterative methods to the convergence criteria
 $\bar \omega$. Tables \ref{table_1a}, \ref{table_1b}, \ref{table_1c}  
correspond respectively to  $\bar \omega$=$10^{-6}$, $10^{-8}$, and $10^{-10}$. 
Number of points per decade in the logarithmic $\tau$ scale
is denoted by [$N_{\rm pts}/D$]. The SOR parameter used is 1.5. The entries
under each method indicate the number of iterations required for convergence. }
\label{table_1}
\subtable[$\bar \omega$=$10^{-6}$]{
\begin{tabular}{crrrrr}
\tableline\tableline
  {[$N_{\rm pts}/D$]} & {Jacobi} & {GS} & {SOR} & {Pre-BiCG} & {Pre-BiCG-STAB} \\
\tableline
 5 & 81  & 40 & 24 & 16 & 12 \\
 8 & 136 & 69 & 22 & 19 & 15 \\
 30 & 444 & 230 & 74 & 33 & 23 \\
\tableline
\end{tabular}
\label{table_1a}
}
\subtable[$\bar \omega$=$10^{-8}$]{
\begin{tabular}{crrrrr}
\tableline\tableline
 {[$N_{\rm pts}/D$]} & {Jacobi} & {GS} & {SOR} & {Pre-BiCG} & {Pre-BiCG-STAB} \\
\tableline
 5 & 110 & 54 & 30 & 18 & 13 \\
 8 & 186 & 94 & 30 & 22 & 15 \\
 30 & 635 & 325 & 103 & 39 & 30\\
\tableline
\end{tabular}
\label{table_1b}
}
\subtable[$\bar \omega$=$10^{-10}$]{
\begin{tabular}{crrrrr}
\tableline\tableline
{[$N_{\rm pts}/D$]} & {Jacobi} & {GS} & {SOR} & {Pre-BiCG}  & {Pre-BiCG-STAB}\\
\tableline
5 & 138 & 68 & 37 & 20  & 14\\
8 & 236 & 118  & 40 & 25 & 18 \\
30 & 827 & 419 & 132 & 45 & 30\\
\tableline
\end{tabular}
\label{table_1c}
}
\end{center}
\end{table*}
%%%%%%%%%%%%%%%%%%%%%%%%%%%%%%%%%%%%%%%%%%%%%%%%%%%%%%%%table
%%%%%%%%%%%%%%%%%%%%%%%%%%%%%table
\begin{table*}
\begin{center}
\caption{Timing efficiency of the iterative methods}
\label{table_2}
\begin{tabular}{crrrrr}
\tableline\tableline
  & {Jacobi} & {GS} & {SOR} & {Pre-BiCG} & {Pre-BiCG-STAB}\\
\tableline
CPU time for convergence & 7 min 49 sec & 4 min 4 sec & 1 min 18 sec & 27 sec 
& 42 sec\\
Overrates in computing &6 sec & 6 sec & 6 sec & 9 sec & 6 sec \\
Total computing time & 7 min 55 sec & 4 min 10 sec & 1 min 24 sec & 36 sec 
& 48 sec \\
\tableline
\end{tabular}
\end{center}
\end{table*}
%%%%%%%%%%%%%%%%%%%%%%%%%%%%%%%%%%%%%%%%%%%%%%%%%%%%table
\appendix
\section{Construction of $\hat A$ matrix and Preconditioner matrix $\hat M$}
\label{appendix_lambda}
In Pre-BiCG method, it is essential to compute and store the $\hat A^T$ matrix.
A brute force - fully numerical way of doing this is as follows. Suppose that
the dimension of $\hat A$ matrix is $N_{\rm d} \times N_{\rm d}$, where 
$N_{\rm d}$ is the number of depth points. By sending a $\delta$-source function
$N_{\rm d}$ times, to a formal solver subroutine, $N_{\rm d}$ columns of 
$\hat A$ matrix can be calculated. But this takes a large amount of CPU time
especially for large values of $N_{\rm d}$.

Instead, there is a semi-analytic way of calculating the $\hat A$ matix. 
By substituting the $\delta$-source function in the expression for the
intensity on a short-characteristic stencil of 3-points (MOP in standard 
notation) we can obtain ``recursive relations for intensity matrix elements
$\bm{I}_{\rm ij} (i,j,=1,2, \ldots N_{\rm d})$, which can then be integrated
over frequencies and angles to get the $\hat \Lambda$ matrix. Finally,
$\hat A=[\hat I - (1-\epsilon) \hat \Lambda]$. The diagonal of 
$\hat \Lambda$ is $\hat \Lambda^*$ and the diagonal of $\hat A$ 
is the preconditioner matrix $\hat M$.

For plane parallel full-slab problem, this is given in \citet{kun88}.
In radiative transfer problems with spherical symmetry, it is sufficient
to compute the solution on a quadrant. However this causes a tricky
situation, in which we have to define a mid-line 
(see Fig.~\ref{fig_geometry}) on which a non-zero boundary 
condition $\bm{I}^+=\bm{I}^-$ has to be specified.
For the outgoing rays, the mid-vertical line is the starting grid point for
a given ray. Since the intensity at the starting point is non-zero 
($\bm{I}^+=\bm{I}^-$), intensity at any interior point 
depends on the intensity
at all the previous points. Recall that
\begin{equation}
\bm{I}_{\rm k} (\mu > 0)= \bm{I}_{\rm k+1} (\mu > 0)
\exp(-\Delta \tau_{\rm k}(\mu >0))+\bm{\Psi}_{\rm k-1} \bm{S}_{\rm k-1}
+\bm{\Psi}_{\rm k} \bm{S}_{\rm k}+\bm{\Psi}_{\rm k+1} \bm{S}_{\rm k+1}
\end{equation}
and
\begin{equation}
\bm{I}_{\rm k+1} (\mu > 0)= \bm{I}_{\rm k+2} (\mu > 0)
\exp(-\Delta \tau_{\rm k+1}(\mu >0))+\bm{\Psi}_{\rm k} \bm{S}_{\rm k}
+\bm{\Psi}_{\rm k+1} \bm{S}_{\rm k+1}+\bm{\Psi}_{\rm k+2} \bm{S}_{\rm k+2}
\end{equation}
and so on until we reach the mid-line. It is easy to see from above
equations that intensity calculation at a short-characteristic stencil MOP is
not confined only to the intensity on MOP, but also on all previous points, 
through spatial coupling. This is specific to performing radiative 
transfer on a spherical quadrant. Note that even for the construction of a
diagonal $\hat \Lambda$, all the elements $\bm{I}_{\rm ij}$ of the 
intensity matrix has to be computed. We present below the recursive 
relations to compute $\bm{I}_{\rm ij} (i,j=1,2,\ldots,N_{\rm d})$.\\

\noindent
{\bf For the incoming rays ($\mu < 0$)  - Reverse sweep}\\
DO $i=1, 2, \ldots , N_{\rm d}$\\

\noindent
Consider an arbitrary spatial point $i$. The delta-source vector is 
specified as\\
\begin{eqnarray}
\bm{S}(\tau_{\rm i}) = 1, \quad \bm{S}(\tau_{\rm j}) = 0 \quad \textrm{for} \quad i \neq j.
\end {eqnarray}
DO $ip=1, 2, \ldots , N_{\rm p}$, where $N_{\rm p}$ is the total number of
impact parameters.\\

\noindent
For the inner boundary points, define $N_{\rm v_{p}} = N_{\rm d}$ for the 
core rays and $N_{\rm v_{p}} = N_{\rm d}-(ip-N_{\rm c}-1)$ for lobe rays.
The index $N_{\rm v_{p}}$ represents the total number of points on a given 
ray of constant impact parameter $p$. The external boundary condition
has to be taken as zero for constructing integral operators like 
$\hat \Lambda$. 
\begin{equation}
\bm{I}_{\rm i1}(\tau_{\rm i},\tau_1, x, p)
= 0.
\end{equation}

\noindent
For those rays (with index $ip$) for which $p(ip) \le r(i)$\\
DO $j=2, 3, \ldots N_{\rm v_{p}}$ \\
IF\\
$(j=i+1)$ and ($j=N_{\rm d}$ or $p(ip)=r(j)$), which are interior boundary 
points
\begin{equation}
\bm{I}_{\rm i,j}(\tau_{\rm i},\tau_{\rm j}, x, p)=
\bm{I}_{\rm i,j-1}(\tau_{\rm i},\tau_{\rm j-1}, x, p) 
\exp{(-\Delta \tau_{\rm j}(\mu))}
+\bm{\Psi}_{\rm u}(j, x, p, \mu < 0) + \bm{\Psi}_{\rm d}(j, x, p, \mu < 0)
\end{equation}

\noindent
This is because at these interior boundary points we assume 
$\bm{S}_{\rm d}=\bm{S}_{\rm u}$ and $\bm{S}_{\rm d}=\bm{S}_{\rm u}=1$
when $j=i+1$.\\

\noindent
ELSE\\
(Non interior boundary points)\\
If $j=i+1$ 
\begin{equation}
\bm{I}_{\rm i,j}(\tau_{\rm i},\tau_{\rm j}, x, p)=
\bm{I}_{\rm i,j-1}(\tau_{\rm i},\tau_{\rm j-1}, x, p) 
\exp{-(\Delta \tau_{\rm j}(\mu))}
+\bm{\Psi}_{\rm d}(j, x, p, \mu < 0) 
\end{equation}

\noindent
Elseif $j=i$
\begin{equation}
\bm{I}_{\rm i,j}(\tau_{\rm i},\tau_{\rm j}, x, p)=
\bm{I}_{\rm i,j-1}(\tau_{\rm i},\tau_{\rm j-1}, x, p) 
\exp{(-\Delta \tau_{\rm j}(\mu))}
+\bm{\Psi}_{\rm 0}(j, x, p, \mu < 0) 
\end{equation}

\noindent
Elseif $j=i-1$
\begin{equation}
\bm{I}_{\rm i,j}(\tau_{\rm i},\tau_{\rm j}, x, p)=
\bm{I}_{\rm i,j-1}(\tau_{\rm i},\tau_{\rm j-1}, x, p) 
\exp{(-\Delta \tau_{\rm j}(\mu))}
+\bm{\Psi}_{\rm u}(j, x, p, \mu < 0) 
\end{equation}

\noindent
Else\\
if ($j \le i-2$)
\begin{equation}
\bm{I}_{\rm i,j}(\tau_{\rm i},\tau_{\rm j}, x, p)=0
\end {equation}
else
\begin{equation}
\bm{I}_{\rm i,j}(\tau_{\rm i},\tau_{\rm j}, x, p)=
\bm{I}_{\rm i,j-1}(\tau_{\rm i},\tau_{\rm j-1}, x, p) 
\exp{(-\Delta \tau_{\rm j}(\mu))}
\end{equation}
end if\\
End if\\
END IF\\
END DO\\

\noindent
For those rays for which $p(ip)>r(i)$\\
DO $j=2, 3, \ldots N_{\rm v_{p}}$ \\
\begin{equation}
\bm{I}_{\rm i,j}(\tau_{\rm i},\tau_{\rm j}, x, p)=0
\end{equation}
END DO\\
END DO\\
END DO\\

\noindent
{\bf For the outgoing rays ($\mu > 0$)  - Forward sweep }\\
Let $i=N_{\rm d}, N_{\rm d}-1, \ldots, 1$\\
\begin{eqnarray}
\bm{S}(\tau_{\rm i}) = 1, \quad \bm{S}(\tau_{\rm j}) = 0 \quad \textrm{for} \quad i \neq j
\end {eqnarray}
DO $ip=1, 2, \ldots , N_{\rm p}$\\
For $j=N_{\rm v_{p}}$ (Inner boundary point) \\
\begin{equation}
\bm{I}_{\rm i,j}(\tau_{\rm i},\tau_{\rm j}, x, p) \quad (\mu > 0) =
\bm{I}_{\rm i,j}(\tau_{\rm i},\tau_{\rm j}, x, p) \quad (\mu < 0)
\end{equation}

\noindent
For non boundary points\\

\noindent
For those rays (with index $ip$) for which $p(ip) \le r(i)$\\
DO $j=N_{\rm v_{p}}-1, N_{\rm v_{p}}-2, \ldots 1$ \\
\noindent
If $j=i+1$ 
\begin{equation}
\bm{I}_{\rm i,j}(\tau_{\rm i},\tau_{\rm j}, x, p)=
\bm{I}_{\rm i,j+1}(\tau_{\rm i},\tau_{\rm j+1}, x, p) 
\exp{(-\Delta \tau_{\rm j}(\mu))}
+\bm{\Psi}_{\rm d}(j, x, p, \mu > 0) 
\end{equation}

\noindent
Elseif $j=i$
\begin{equation}
\bm{I}_{\rm i,j}(\tau_{\rm i},\tau_{\rm j}, x, p)=
\bm{I}_{\rm i,j+1}(\tau_{\rm i},\tau_{\rm j+1}, x, p) 
\exp{(-\Delta \tau_{\rm j}(\mu))}
+\bm{\Psi}_{\rm 0}(j, x, p, \mu > 0) 
\end{equation}

\noindent
Elseif $j=i-1$
\begin{equation}
\bm{I}_{\rm i,j}(\tau_{\rm i},\tau_{\rm j}, x, p)=
\bm{I}_{\rm i,j+1}(\tau_{\rm i},\tau_{\rm j+1}, x, p) 
\exp{(-\Delta \tau_{\rm j}(\mu))}
+\bm{\Psi}_{\rm u}(j, x, p, \mu > 0) 
\end{equation}

\noindent
Else
\begin{equation}
\bm{I}_{\rm i,j}(\tau_{\rm i},\tau_{\rm j}, x, p)=
\bm{I}_{\rm i,j+1}(\tau_{\rm i},\tau_{\rm j+1}, x, p) 
\exp{(-\Delta \tau_{\rm j}(\mu))}
\end{equation}
End if\\
END DO\\

\noindent
For those rays for which $p(ip)>r(i)$\\
DO $j=N_{\rm v_{p}}, N_{\rm v_{p}}-1, \ldots, 1$ \\
\begin{equation}
\bm{I}_{\rm i,j}(\tau_{\rm i},\tau_{\rm j}, x, p)=0
\end{equation}
END DO\\
END DO\\
END DO\\

\noindent
The algorithm given above saves a great deal of computing time
by cutting down the number of calls to the formal solver -2 instead
of $N_{\rm d}$ -the first call to store the $\bm{\Psi}$ and $\Delta \tau$
at all depth points, and the second call to compute $\bm{I}_{\rm ij}$


\begin{thebibliography}{}
\bibitem[Asensio Ramos \& Trujillo Bueno(2006)]{aar06}
Asensio Ramos, A., \& Trujillo Bueno, J. 2006, in EAS Pub. Ser. 18, 
Radiative Transfer and Applications to Very Large Telescopes, ed. 
Ph. Stee, 25
\bibitem[Auer(1984)]{aue84}
Auer, L.~H. 1984, in  Methods in radiative transfer, ed.
Kalkofen, W. (Cambridge: Cambridge University Press), 237
\bibitem[Auer(1984)]{aue91}
Auer, L.~H. 1991, in  Stellar Atmospheres: Beyond Classical Models, ed.
Crivellari,~L., Hubeny,~I., \& Hummer,~D.~G. (Dordrecht: Kluwer Academic
Publishers), 9
\bibitem[Auer et al.(1994)]{aue94}
Auer, L.~H., Fabiani Bendicho, P., \& Trujillo Bueno, J. 1994, \aap, 292, 599
\bibitem[Cannon(1973)]{can73}
Cannon, C. J. 1973, JQSRT, 13, 627
\bibitem[Chandrasekhar(1934)]{cha34}
Chandrasekhar, S. 1934, MNRAS, 94, 522
\bibitem[Chevallier et al.(2003)]{che03}
Chevallier, L., Paletou, F., \& Rutily, B. 2003, \aap, 411, 221
\bibitem[Daniel \& Cernicharo(2008)]{dan08}
Daniel, F., \& Cernicharo, J. 2008, \aap, 488, 1237
\bibitem[Gros et al.(1997)]{gro97}
Gros, M., Crivellari, L., \& Simonneau, E. 1997, \apj, 489, 331
\bibitem[Hamann(1985)]{hum85}
Hamann, W-R. 1985, \aap, 145, 443
\bibitem[Hamann(2003)]{hum03}
Hamann, W-R. 2003, in ASP Conf. Ser. 288, Stellar Atmosphere Modeling, ed.
Hubeny, I., Mihalas, D., \& Werner, K. (San Francisco: ASP), 171
\bibitem[Hestenes \& Stiefel(1952)]{hes52}
Hestenes, M.~R., \& Stiefel, E. 1952, Journal of Research of the National
Bureau of Standards, 49(6), 409
\bibitem[Hubeny \& Burrows (2007)]{hub07}
Hubeny, I., \& Burrows, A. 2007, \apj, 659, 1458
\bibitem[Hubeny(2003)]{hub03}
Hubeny, I. in ASP Conf. Ser. 288, Stellar Atmosphere Modeling, ed.
Hubeny, I., Mihalas, D., \& Werner, K. (San Francisco: ASP), 17
\bibitem[Hummer \& Rybicki(1971)]{hum71}
Hummer, D.~G., \& Rybicki, G.~B. 1971, MNRAS, 152, 1
\bibitem[Klein et al.(1989)]{kle89}
Klein, R.~I., Castor, J.~I., Greenbaum, A., Taylor, D., \&
Dykema, P.~G. 1989, JQSRT, 41, 199
\bibitem[Kosirev(1934)]{kos34}
Kosirev, N.~A. 1934, MNRAS, 94, 430
\bibitem[Kunasz \& Auer(1988)]{kun88}
Kunasz, P.~B., \& Auer, L.~H. 1988, JQSRT, 39, 67
\bibitem[Kunasz \& Hummer(1973)]{kun73}
Kunasz, P.~B., \& Hummer, D.~G. 1973, MNRAS, 166, 57
\bibitem[Kunasz \& Hummer(1974)]{kun74}
Kunasz, P.~B., \& Hummer, D.~G. 1974, MNRAS, 166, 19
\bibitem[Mihalas(1978)]{mih78}
Mihalas, D. 1978, Stellar Atmospheres (2nd ed.; San Francisco: Freeman)
\bibitem[Olson et al.(1986)]{ols86}
Olson, G.~L., Auer, L.~H., \& Buchler, J.~R. 1986, JQSRT, 35, 431
\bibitem[Paletou \& Anterrieu(2009)]{pal09}
Paletou, F., \& Anterrieu. 2009, arXiv:0905.3258, 
http://arxiv.org/abs/0905.3258
\bibitem[Peraiah(2002)]{per02}
Peraiah, A. 2002, An Introduction to Radiative Transfer (Cambridge University Press)
\bibitem[Peraiah \& Grant(1973)]{per73}
Peraiah, A., \& Grant, I. P. 1973, JIMA, 12, 75
\bibitem[Saad(2000)]{saad00}
Saad, Y. 2000, Iterative methods for Sparse Linear Systems (2nd ed.)
\bibitem[Scharmer(1981)]{sch81}
Scharmer, G. B. 1981, \apj, 249, 720 
\bibitem[Schmid-Burgk(1974)]{sch74}
Schmid-Burgk, J. 1974, \aap, 32, 73
\bibitem[Trujillo Bueno \& Fabiani Bendicho(1995)]{tru95}
Trujillo Bueno, J., \& Fabiani Bendicho, P. 1995, \apj, 455, 646
\bibitem[Werner \& Husfeld(1985)]{wer85}
Werner, K., \& Husfeld, D. 1985, \aap, 148, 417

\end{thebibliography}
\end{document}